 \documentclass[iop]{emulateapj}

\usepackage{rotating, afterpage}

\usepackage{graphicx}
\usepackage{amsmath}

\slugcomment{Prepared for the Astrophysical Journal}

\shorttitle{The Formation of the Wide Asynchronous Binary Asteroid Population}
\shortauthors{Jacobson, Scheeres \& McMahon}

\begin{document}

\title{The Formation of the Wide Asynchronous Binary Asteroid Population}

\author{Seth A. Jacobson\altaffilmark{1,2,3}}
\affil{Department of Astrophysical and Planetary Science, UCB 391, University of Colorado, Boulder, CO 80309, USA}

\author{Daniel J. Scheeres\altaffilmark{1}}
\affil{Department of Aerospace Engineering Sciences, UCB 429, University of Colorado, Boulder, CO 80309, USA}

\and

\author{Jay McMahon\altaffilmark{1}}
\affil{Department of Aerospace Engineering Sciences, UCB 429, University of Colorado, Boulder, CO 80309, USA}

\altaffiltext{1}{Colorado Center for Astrodynamics Research, UCB 431, University of Colorado, Boulder, CO 80309}
\altaffiltext{2}{Laboratoire Lagrange, Observatoire de la C\^{o}te d'Azur, Nice 06304 Cedex 4, France}
\altaffiltext{3}{Bayerisches Geoinstitut, Universit\"{a}t Bayreuth, 95440 Bayreuth, Germany}

\begin{abstract}
We propose and analyze a new mechanism for the formation of the wide asynchronous binary population. These binary asteroids have wide semi-major axes relative to most near-Earth and Main Belt asteroid systems. Confirmed members have rapidly rotating primaries and satellites that are not tidally locked. Previously suggested formation mechanisms from impact ejecta, planetary flybys and directly from rotational fission events cannot satisfy all of the observations. The newly hypothesized mechanism works as follows: (i) these systems are formed from rotational fission, (ii) their satellites are tidally locked, (iii) their orbits are expanded by the BYORP effect, (iv) their satellites de-synchronize due to the adiabatic invariance between the libration of the secondary and the mutual orbit, and (v) the secondary avoids resynchronization due to the the YORP effect. This seemingly complex chain of events is a natural pathway for binaries with satellites that have particular shapes, which define the BYORP effect torque that acts on the system. After detailing the theory, we analyze each of the wide asynchronous binary members and candidates to assess their most likely formation mechanism. Finally, we suggest possible future observations to check and constrain our hypothesis.
\end{abstract}


\section{Introduction}
The observation that most small near-Earth and Main Belt asteroid binary systems have a rapidly rotating primary is one of the key pieces of evidence that led astronomers to more closely investigate rotational fission as the dominant binary formation mechanism~\citep{Margot:2002fe}.~\citet{Scheeres:2007io} and~\citet{Walsh:2008gk} showed that the creation of binary asteroid systems is possible via YORP-induced rotational fission.~\citet{Jacobson:2011eq} modeled the dynamics of rotationally fissioned asteroids and determined the properties of these newly created binaries, which include the primary\footnote{The primary is always the more massive binary member. The secondary is the less massive.} always rotating rapidly prograde compared to the mutual orbit rate and the secondary rotating at a different rate than the mutual orbit (typically faster). As observed in numerical experiment, the rotational fission formation mechanism creates tight binary systems with a median mutual semi-major axis of $a = 3.3$ $R_p$ and a maximum of $17$ $R_p$~\citep{Jacobson:2011eq}. 

After creation, tidal synchronization of the secondary is the fastest tidal process within the binary system~\citep{Goldreich:2009ii}. For low mass ratio systems\footnote{The mass ratio $q$ is the secondary mass divided by the primary mass.} $q \lesssim 0.2$ synchronization of the primary can take more than an order of magnitude longer than the secondary, and for given estimates of the relevant tidal parameters, tidal synchronization of the secondary is also shorter than the dynamical or collisional lifetime of kilometer and sub-kilometer binary systems for the near-Earth or Main Belt asteroid populations, respectively. The outcome of this tidal process is a tidally-locked secondary orbiting a still rapidly rotating primary. These singly synchronous\footnote{Singly synchronous refers to the synchronicity of the secondary spin period and the mutual orbit period. Some authors in the past have referred to these systems as asynchronous referring to the lack of synchronicity of the primary period and the mutual orbit period. We reserve the label asynchronous for those systems that have absolutely no synchronicity within the system.} binary systems are the most prevalent small ($R_p < 10$ km) binary asteroid systems in either the near-Earth or Main Belt asteroid populations~\citep{Margot:2002fe,Pravec:2006bc}. 

Two effects, mutual body tides and the binary YORP (BYORP) effect, continue to evolve singly synchronous binaries. Each is described in detail in Section~\ref{sec:TidalandBYORPevolution}. According to the tidal-BYORP equilibrium hypothesis proposed in~\citet{Jacobson:2011hp}, singly synchronous binary systems can evolve to a long-term stable semi-major axis equilibrium if the torque from the BYORP effect on the mutual orbit is contractive (i.e. acting opposite to the always expansive tidal torque). However, if the BYORP torque is expansive, then the two torques grow the mutual orbit to the Hill radius, where the orbit is disrupted by distant third body encounters~\citep{Cuk:2007gr,McMahon:2010jy,Jacobson:2011hp}. 

Herein, we propose that these expanding singly synchronous binary systems are also the source of the wide asynchronous binary asteroid population, which is described in detail in Section~\ref{sec:thewideasynchronousbinaryasteroidpopulation}. There exists an adiabatic invariance between the size of the mutual orbit and the libration amplitude of the tidally locked secondary, and so as the mutual orbits of these systems expand, the libration amplitude increases. At some large semi-major axis, the libration amplitude reaches $90^\circ$ and the secondary begins to circulate. This circulation turns off the BYORP effect, which at large binary separations, is the only significant torque acting on the orbit, since tidal evolution is very weak at these distances. Once the secondary has begun circulating, it is rotationally accelerated by the YORP effect. This process is similar to YORP acceleration of a single body~\citep{Bottke:2006en}, since tides are so weak.

Since the strength of mutual body tides is inversely proportional to the semi-major axis to the sixth power~\citep{Goldreich:2009ii}, tides are much less effective on wider binaries than they are on the more frequently observed tight binary population. For instance, consider two identical systems with the exception of their semi-major axes. The first has the median semi-major axis $a = 3.3$ $R_p$ found for post-fission binary systems~\citep{Jacobson:2011eq}, and the second has a semi-major axis of $a = 12$ $R_p$. The second, then takes $\sim 10^3$ times longer to tidally damp than the first. This damping timescale ratio is the same for the synchronization of each body and the circularization of the orbit. Considering that the eccentricity damping timescales are thought to be between $\sim10^4$ and $\sim10^7$ years for tight binary systems ($2 < a < 8$ $R_p$)~\citep{Fang:2012fw}, then the tightest wide asynchronous candidates ($a = 12$ $R_p$) have damping timescales between $10^7$ and $10^{10}$ years.

\begin{figure}[tb!]
\begin{center}
\includegraphics[width=\columnwidth]{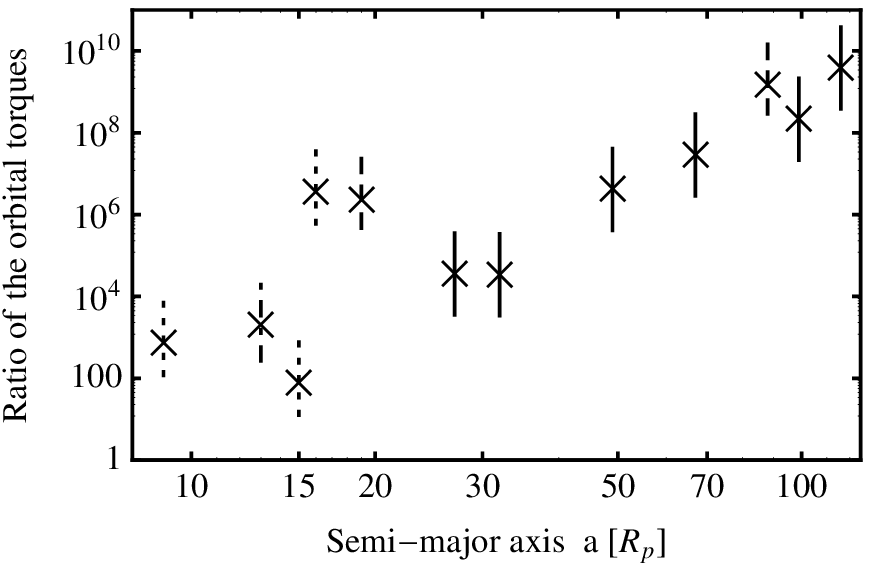}
\includegraphics[width=\columnwidth]{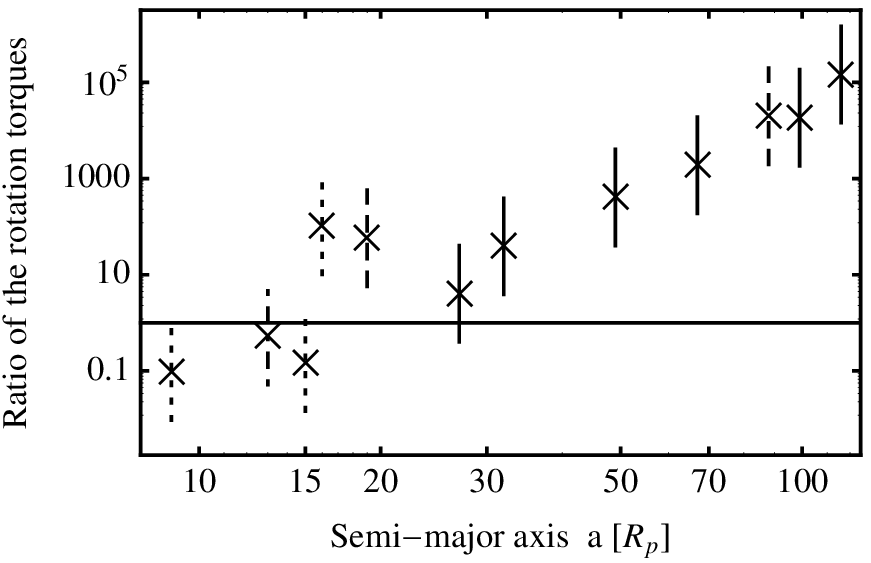}
\end{center}
\caption{The top figure shows the ratio of the orbital torque from the BYORP effect to the tidal torque on the orbit from the tidal bulge raised on the primary by the secondary as a function of the semi-major axis of each wide asynchronous candidate system and wide asynchronous pair in each triple system. The bottom figure shows the ratio of the YORP effect rotation torque and the tidal torque on the rotation of the secondary from the primary as a function of the semi-major axis of the same systems as the top figure. The horizontal line is at $1$, where the two torques are equal. The equations for these torques are found in Appendix~\ref{sec:derivationofthetidaltorqueratiosinFigure1} and the data for each system is listed in Tables~\ref{tab:binaryparameters} and~\ref{tab:tripleparameters}. Uncertainty is dominated by the poorly known BYORP, YORP and tidal parameters. From left to right the systems are 51356 (2000 RY$_{76}$), 153591 (2001 SN$_{263}$), 1717 Arlon, 1998 ST$_{27}$, 136617 (1999 CC), 317 Roxane, 32039 (2000 JO$_{23}$), 1509 Esclangona, 22899 (1999 TO$_{14}$), 3749 Balam, 17246 (2000 GL$_{74}$) and  4674 Pauling. The lines are coded such that the three tightest wide binary systems (51356, Arlon and 1998 ST$_{27}$) are dotted lines, two triples systems are dashed (Balam and 136617) while 153591 is dot-dashed since it is a tight triple system, and the rest are solid black lines.}
\label{fig:ratioplots}
\end{figure}
Figure~\ref{fig:ratioplots} further motivates the identification of this particular class of small binary, the wide asynchronous binary, and the expanding singly synchronous formation mechanism described above. It shows that the radiative BYORP and YORP effects are much more important than the mutual body tidal torques for the wide asynchronous binaries. A handful of wide asynchronous candidates may have secondaries with tidal torques stronger than the YORP effect. These systems will be discussed in Section~\ref{sec:individualsystems} in the context of alternative formation mechanisms. The clear trend of increasing ratio number with semi-major axis affirms that the tidal dependence on separation distance is the most important factor.

The wide asynchronous binary populations are very different than the singly synchronous asteroids, which as~\citet{Jacobson:2011hp} proposed are a result of tides and the BYORP effect having equivalent strengths but opposite directions. However, by considering that the BYORP effect and tides can act in unison rather than in opposition, it is possible to connect these two distinct populations. The key to the proposed expanding singly synchronous hypothesis is the adiabatic invariance, which transforms outward orbit expansion into excited libration of the secondary. Since the BYORP effect drives the orbital expansion and the BYORP effect requires a synchronized (even if librating) secondary, the widening of these systems becomes the very act that strands them at wide separations. This mechanism for the creation of wide asynchronous binaries requires specific initial conditions, namely a wider than typical orbit after rotational fission (as can be seen in Figure~\ref{fig:xplot}). Since not all expanding singly synchronous binaries will start at these initial separations, some systems will expand to their Hill radii and become asteroid pairs. The future observational determination of the ratio of tight to wide binaries would constrain just how sensitive these initial conditions are.

It is important to note that not all asynchronous binary systems are wide. Tight asynchronous binary systems have semi-major axes $a \leq 8$ $R_p$ which is also the semi-major axis cut-off observed in the singly synchronous binary population. This tight population is consistent with either recent formation from rotational fission (i.e. synchronization of the secondary is currently underway) or strong planetary perturbations, which essentially re-set the system and require the secondary to be synchronized again to the new, typically slower orbital period~\citep{Jacobson:2011eq,Fang:2012go,Jacobson:2012tx}. 

\subsection{The wide asynchronous binary asteroid population}
\label{sec:thewideasynchronousbinaryasteroidpopulation}
The population of wide asynchronous binary systems includes four known members and five suspects. All candidates\footnote{For convenience, when we want to refer to all of them we'll call them all candidates since we're considering them for this possible creation mechanism.} have semi-major axes $a > 8$ primary radii\footnote{When comparing binary systems to one another, it is important to use properly scaled measurements. Here we are comparing the semi-major axes of the system measured in primary radii.} $R_p$, larger than any member of the singly synchronous population (e.g. 66391 (1999 KW$_4$)), which have semi-major axes $a < 8$ $R_p$ and are distributed about a median of $a = 4$ $R_p$. Wide asynchronous binary systems are distinct from other wide binary systems that have doubly synchronous members and high mass ratios (e.g. 4951 Iwamoto (1990 BM)), since all wide asynchronous candidates have low mass ratios and at least one asynchronous member. Possessing a second asynchronous member is the distinction between being a suspect and full membership in the wide asynchronous class. These wide asynchronous binaries are also distinct from the tight asynchronous binaries (e.g. 35107 (1991 VH)), which have a similar distribution of semi-major axes as the singly synchronous population with the same observed upper limit of $8$ $R_p$. All known wide asynchronous members have a $R_p \lesssim 10 $ km. The YORP effect is not efficient at driving asteroids to rotational fission above this size due to competition with collisional evolution and the dynamical lifetime of the system. We limit the suspected members to this same size range, and note that the other binary asteroid classes hypothesized to form from YORP-induced rotational fission obey this size cut-off. To summarize, wide asynchronous binary candidates have low mass ratios, small sizes and rapidly rotating primaries with the exception of 317 Roxane and possibly 1717 Arlon. These characteristics are identical to the singly synchronous binary population~\citep{Jacobson:2011hp}, but all of these systems have larger semi-major axes and those with measured secondary rotation periods are asynchronous. Statistics regarding binaries in this paragraph are from the July 1, 2011 binary asteroid parameter release found at \url{http://www.asu.cas.cz/asteroid/binastdata.htm}. It is compiled by methods described in~\citet{Pravec:2007fw} and maintained by Petr Pravec and collaborators.

\begin{deluxetable*}{ l c c | c | c | cc | cc | cc }
\tablecolumns{11}
\tabletypesize{\scriptsize}
\tablewidth{\textwidth}
\tablecaption{Wide asynchronous binary asteroids.\label{tab:binaryparameters}}
\tablehead{\colhead{Asteroid System} & \colhead{$a_\odot$ [AU]} & \colhead{$e_\odot$} &  \colhead{$q$} & \colhead{$R_p$ [km]} & \colhead{$a$ [$R_p$]} & \colhead{$a$ [km]} & \colhead{$P_p$ [$P_d$]} & \colhead{$P_p$ [h]} & \colhead{$P_s$ [$P_d$]} & \colhead{$P_s$ [h]} }
\startdata
317 Roxane\tablenotemark{A} & $2.29$ & $0.09$ & $0.023$\tablenotemark{B} & $9.4$\tablenotemark{B} & $27$ & $257$\tablenotemark{B} & $3.5$ & $8.2$\tablenotemark{C} & \nodata & \nodata \\
1509 Esclangona (1938 YG)\tablenotemark{D} & $1.87$ & $0.03$ & $0.036$\tablenotemark{D} & $4.3\tablenotemark{E}$ & $49$ & $210$\tablenotemark{D} & $1.4$ & $3.3$\tablenotemark{F} & $2.8$ & $6.6$\tablenotemark{F} \\
1717 Arlon (1954 AC)\tablenotemark{G} & $2.20$ & $0.13$ & $>0.216$\tablenotemark{G} & $3.9$\tablenotemark{H} & $15$ & $59$\tablenotemark{H} & $2.2$ & $5.1$\tablenotemark{G,*} & $7.8$ & $18.2$\tablenotemark{G,*}  \\
4674 Pauling (1989 JC)\tablenotemark{J} & $1.86$ & $0.07$ & $0.033$\tablenotemark{J}  & $2.2$\tablenotemark{K} & $116$ & $250$\tablenotemark{J} & $1.1$ & $2.5\tablenotemark{L}$ & \nodata & \nodata \\
17246 (2000 GL$_{74}$)\tablenotemark{M} & $2.84$ & $0.02$ & $0.064$\tablenotemark{M} & $2.3$\tablenotemark{H} & $99$ & $228$\tablenotemark{B} &  \nodata & \nodata & \nodata & \nodata \\
22899 (1999 TO$_{14}$)\tablenotemark{N} & $2.84$ & $0.08$ & $0.033$\tablenotemark{N} & $2.7$\tablenotemark{O} & $67$ & $182$\tablenotemark{B} & $1.7$ & $4.0$\tablenotemark{P} & \nodata & \nodata \\
32039 (2000 JO$_{23}$)\tablenotemark{Q} & $2.22$ & $0.28$ & $ 0.275$\tablenotemark{Q} & $1.3$\tablenotemark{H} & $32$ & $42$\tablenotemark{H} & $1.4$ & $3.3$\tablenotemark{Q} & $4.8$ & $11.1$\tablenotemark{Q}  \\
51356 (2000 RY$_{76}$)\tablenotemark{R} & $1.81$ & $0.11$ & $0.009$\tablenotemark{R} & $1.2$\tablenotemark{H} & $9$ & $11$\tablenotemark{H} & $1.1$ & $2.6$\tablenotemark{R} & \nodata & \nodata \\
1998 ST$_{27}$\tablenotemark{S} & $0.82$ & $0.53$ & $0.0034$\tablenotemark{S} & $0.28$\tablenotemark{S} & $16$ & $4.5$\tablenotemark{T} & $1.3$ & $3.1$\tablenotemark{S} & $2.6$ & $\lesssim6.0$\tablenotemark{S} 
\enddata
\tablenotetext{A}{\citet{Merline:2009vs}}
\tablenotetext{B}{\citet{Durda:2010vz}} 
\tablenotetext{C}{\citet{Harris:1992ki}} 
\tablenotetext{D}{\citet{Merline:2003ut}}
\tablenotetext{E}{\citet{Marchis:2012ft}}
\tablenotetext{F}{\citet{Warner:2010vb}} 
\tablenotetext{G}{\citet{Cooney:2006ur}}
\tablenotetext{H}{Calculated values consistent with reported observations determined by Petr Pravec and colleagues and reported in the Binary Asteroid Parameters data found at \url{http://www.asu.cas.cz/asteroid/binastdata.htm}, which is compiled by methods described in~\citet{Pravec:2007fw}.}
\tablenotetext{J}{\citet{Merline:2004vj}}
\tablenotetext{K}{\citet{Pravec:2012hs}}
\tablenotetext{L}{\citet{Warner:2006wj}}
\tablenotetext{M}{\citet{Tamblyn:2004vo}}
\tablenotetext{N}{\citet{Merline:2003tu}}
\tablenotetext{O}{\citet{Masiero:2011jc}}
\tablenotetext{P}{\citet{Polishook:2011if}}
\tablenotetext{Q}{\citet{Pray:2007td}}
\tablenotetext{R}{\citet{Warner:2013wg}}
\tablenotetext{S}{\citet{Benner:2003ub}}
\tablenotetext{T}{\citet{Brozovic:2011ib}}
\tablecomments{Columns are the heliocentric semi-major axis $a_\odot$ and eccentricity $e_\odot$,  mass ratio $q$, radius of the primary $R_p$, mutual semi-major axis $a$ measured in both primary radii $R_p$ and km, and rotation periods of the primary and the secondary measured in hours and the surface disruption period limit $P_d = \sqrt{3 \pi / \rho G}$, where $\rho$ is the density and $G$ is the gravitational constant. Osculating orbital elements are from JPL Horizons. The published one-sigma uncertainties are often below the precision reported in the table above, but separation distances are sometimes projected distances from adaptive optics direct imaging and not true semi-major axes (please examine the original sources). $^*$In the original report by~\citet{Cooney:2006ur}, it is not clear which period belongs to which body of 1717 Arlon. We have assigned the periods based on the common pattern of a more rapidly rotating primary on the basis of likely formation by rotational fission, however Petr Pravec and colleagues (personal communication) are preparing to report that the periods may more likely belong to the other body as shown above. We also call attention to the large mass ratio of 1717 Arlon, which is only a lower limit. This binary may be very strange indeed.}
\end{deluxetable*}
The names and properties of each wide asynchronous candidate are listed in Table~\ref{tab:binaryparameters}. The next two columns are the heliocentric semi-major axis $a_\odot$ and eccentricity $e_\odot$, which are relevant for radiative torques. The orbits of these systems are very diverse, and include near-Earth, Mars-crossing and Main Belt asteroids. The mass ratio $q$ is in column three. With the possible exception of 1717 Arlon which could be significantly larger, all systems have mass ratios below or near the low mass ratio limit of $q \sim 0.2$ as defined in~\citet{Jacobson:2011eq}. They determined that with all other things being equal low mass ratio binaries tidally synchronize their secondary members much faster than their primary members\footnote{Roughly following the rule $\tau_{p,synch} = \tau_{s,synch} q^{-7/3} $ using the tidal-BYORP equilibrium and classical tidal theory.}. The primary radius $R_p$ is in column four. For many calculations, we use a spherical approximation for both bodies so the secondary radius is $R_s = q^{1/3} R_p$. Then the mutual semi-major axis $a$ is given in both primary $R_p$ and kilometers. For all the equations in the text $a$ has been normalized to the primary radius. This normalization allows us to consider all the systems simultaneously. The next two sets of columns contain the primary period $P_p$ and the secondary period $P_s$ in two different units: $P_d= 2 \pi / \omega_d = \sqrt{3 \pi / \rho G}$, which is the period surface disruption limit, and hours. If a period is near 1 $P_d$ then it is spinning near its surface disruption spin limit. 

\subsection{Outer members of the triple systems}
\begin{deluxetable*}{l c c | c | c | cc | cc | cc }
\tabletypesize{\scriptsize}
\tablecolumns{11}
\tablewidth{\textwidth}
\tablecaption{Outer Members of Triple Systems.\label{tab:tripleparameters}}
\tablehead{
\colhead{Asteroid System} & \colhead{$a_\odot$ [AU]} & \colhead{$e_\odot$} &  \colhead{$q$} & \colhead{$R_p$ [km]} & \colhead{$a$ [$R_p$]} & \colhead{$a$ [km]} & \colhead{$P_p$ [$P_d$]} & \colhead{$P_p$ [h]} & \colhead{$P_t$ [$P_d$]} & \colhead{$P_t$ [h]} }
\startdata
3749 Balam (1982 BG$_1$)\tablenotemark{A} & $2.24$ & $0.11$ & $0.009$\tablenotemark{B} & $3.3$\tablenotemark{B} & $88$ & $289$\tablenotemark{B} & $1.2$ & $2.8$\tablenotemark{C} & \nodata & \nodata \\
136617 (1994 CC)\tablenotemark{D} & $1.64$ & $0.42$ & $0.0035$\tablenotemark{E} & $0.32$\tablenotemark{D} & $19$ & $6.1$\tablenotemark{E} &  $1.0$ &  $2.4$\tablenotemark{D} &  $6.0$ &  $14$\tablenotemark{F} \\
153591 (2001 SN$_{263}$)\tablenotemark{G} & $1.99$ & $0.48$ & $0.026$\tablenotemark{E} & $1.3$\tablenotemark{G} & $13$ & $16.6$\tablenotemark{E} & $1.5$ & $3.4$\tablenotemark{H}  & $5.6$ & $13$\tablenotemark{F}
\enddata
\tablenotetext{A}{\citet{Merline:2002vl}}
\tablenotetext{B}{\citet{Marchis:2008eq}}
\tablenotetext{C}{\citet{Marchis:2008tk}}
\tablenotetext{D}{\citet{Brozovic:2009ib}}
\tablenotetext{E}{\citet{Fang:2011br}}
\tablenotetext{F}{\citet{Brozovic:2011ib}}
\tablenotetext{G}{\citet{Nolan:2008ty}}
\tablenotetext{H}{\citet{Oey:2009vv}}
\tablecomments{Columns are the heliocentric semi-major axis $a_\odot$ and eccentricity $e_\odot$,  mass ratio $q$, radius of the primary $R_p$, mutual semi-major axis $a$ measured in both primary radii $R_p$ and km, and rotation periods of the primary and the ternary measured in hours and the surface disruption period limit $P_d = \sqrt{3 \pi / \rho G}$, where $\rho$ is the density and $G$ is the gravitational constant. Osculating orbital elements are from JPL Horizons. The mass ratio $q$ is the mass of the tertiary (outermost member) divided by the mass of the primary.}
\end{deluxetable*} 

There is another class of wide asynchronous satellites, and these are tertiary (outermost members) of triple asteroid systems. They are listed in Table~\ref{tab:tripleparameters}. With the exception of an interior third body, they resemble the wide asynchronous binaries in every way: mass ratio, absolute size, semi-major axis and their rotation periods. The mass of the interior satellite can be larger or smaller than the exterior satellite whose properties are listed in the table. The interaction with that third member can substantially change the dynamics of the system but in this work, we do not consider that third member.

Forming a triple asteroid system is obviously more complex than forming a binary system, but all three of these systems have rapidly rotating primaries and that is strongly suggestive of rotational fission~\citep{Fang:2011br}. Both satellites could form at the same time from a secondary fission event~\citep{Jacobson:2011eq} or one satellite could form and if it ends up in a large orbit, the primary could go through another YORP-induced rotational fission event forming a stable binary interior to the original binary. Regardless of formation mechanism, the outer binary pair appears similar to the wide asynchronous population, and we will consider these systems as if they evolved similar to the wide asynchronous population by the expanding singly synchronous hypothesis and just ignore the role of the interior ternary member and therefore all possible perturbations such as resonances.

\section{Forming the wide asynchronous population}
Singly synchronous binary asteroid systems are the most numerous observed binary systems~\citep{Pravec:2007fw}. If they occupy the hypothesized tidal-BYORP equilibrium~\citep{Jacobson:2011hp}, then they represent only half of all binary systems which have undergone tidal synchronization, since the direction of the BYORP torque on the secondary is nominally independent of tidal synchronization. We ask, what happens to those systems for which the BYORP effect is expansive? From previous work, we know that those systems with expanding BYORP and tidal torques quickly evolve outward~\citep{Goldreich:2009ii,Cuk:2010im,McMahon:2010jy}, but until now there has not been a comprehensive model of this evolution, which considers both mechanisms and the role of the adiabatic invariance between the size of the mutual orbit and the libration of the secondary.

\subsection{Tidal and BYORP Evolution}
\label{sec:TidalandBYORPevolution}
In the next section, we describe the outward mutual orbit evolution due to both tides and the BYORP effect. It is important to note that throughout we assume that the orbit and spin poles are parallel. This assumption is in good agreement with observation~\citep{Pravec:2012fa}, and it is consistent with formation from a process such as rotational fission that requires a large amount of angular momentum.

For singly synchronous systems, mutual body tides expand the semi-major axis. As the primary rotates beneath the secondary at a rate different than the orbital rate, the primary deforms attempting to obtain a figure in equilibrium with the ever changing gravitational potentials of both bodies. Assuming the primary is not made of perfectly elastic material, this deformation dissipates the spin energy of the primary slowing it down, and the disturbed primary figure transfers angular momentum to the orbit. This transferred angular momentum expands the orbit. If the bodies are in an eccentric orbit, there are similar deformations of each body due to both the radial oscillations of the orbit and the difference between the eccentric and mean anomalies. Mutual body tides are parameterized by the tidal Love number $k$ and the tidal quality number $Q$, which respectively quantify the deformation and the dissipation of the body undergoing the tidal stresses~\citep{Murray:2000th}. The tidal Love number of a perfectly rigid, non-deformable body is $k = 0$ and a body with no dissipation has an infinite tidal quality number. Both parameters are poorly known, but it has be shown that for binary systems with rubble pile components, the combined effects of these tides are to damp the mutual orbit eccentricity and grow the semi-major axis of the system~\citep{Goldreich:2009ii, Jacobson:2011hp}.

The BYORP effect is an averaged radiative torque on the mutual orbit that acts on any system with a body in a spin-orbit resonance~\citep{Cuk:2005hb}. Photons exchange momentum with the body as they are absorbed or emitted. When this change in momentum acts as a torque with a lever arm from the surface point of interaction to the barycenter of the binary asteroid system, it can perturb the mutual orbit. In order for the summation of every small momentum change from each photon to be non-zero, the surface of the body must be in a repeating relationship with the binary barycenter (i.e. spin-orbit resonance) or else they sum to zero over large times. The BYORP effect is the averaged outcome of all of these instantaneous torques. It can expand or contract the mutual orbit depending on the shape and orientation of the body. This is quantified to first order in eccentricity by the BYORP coefficient $B$~\citep{McMahon:2010by}. A BYORP coefficient of $B = 0$ is a completely symmetric body. We define the coefficient so a positive BYORP coefficient corresponds to an expanding mutual orbit $B > 0$.~\citet{McMahon:2010jy} determined that the BYORP effect can double or halve the semi-major axis of a singly synchronous binary system such as 1999 KW$_4$ in $\sim3 - 6 \times 10^{4}$ years. The BYORP effect is less effective with larger libration angles, but it still operates in the same direction.

\citet{Jacobson:2011hp} hypothesized that the prevalence of singly synchronous systems is due to an equilibrium in semi-major axis between the BYORP effect and mutual body tides. In this case, tides expand the system and the BYORP effect contracts it  (i.e. negative BYORP coefficient $B < 0$) driving the mutual orbit to an equilibrium between the two torques. Without exogenous interference, these systems are stable since tides damp eccentricity more strongly than the BYORP effect can excite it. The BYORP coefficient can also be positive $B > 0$ with assumedly the same range of absolute values as the equilibrium systems in~\citet{Jacobson:2011hp} and estimated from shape models by~\citet{McMahon:2012ti}. For example, if the tidally locked secondary in any of the singly synchronous binary systems is flipped so that its near- and far-faces are reversed, then the BYORP coefficient would be of the same strength but opposite sign and so the system would expand. 

When the BYORP torque is expansive, it and mutual body tides damp eccentricity $e$. The time evolution of the eccentricity is thus the linear combination of the tidal evolution~\citep{Goldreich:2009ii} and the BYORP effect evolution~\citep{McMahon:2010jy}: 
\begin{equation}
\dot{e} = - \frac{k_p}{Q} \left( A_T L + a^7 \frac{A_B}{4}  \right) e a^{-13/2}
\label{eqn:eccentricity}
\end{equation}
where $k_p$ is the tidal Love number of the primary, $Q$ is the tidal quality number, $a$ is the semi-major axis measured in primary radii $R_p$, and $A_T$, $L$ and $A_B$ are positive coefficients defined in Appendix~\ref{sec:parameterizationoftheevolutionequations} that depend on various properties of the binary system. Since both torques damp eccentricity, we expect systems evolving outward to have smaller eccentricities. This is in stark contrast to the very high eccentricities expected when a binary forms directly from rotational fission, the expected prevalence of high eccentricity EEBs, or excitation from planetary flybys. However, this does not mean we expect expanding singly synchronous systems and the wide asynchronous binaries they produce to have zero eccentricity, sometimes these systems can expand outwards and desynchronize faster than tides can completely damp the eccentricity. It is a challenge to evaluate this exactly due to the uncertainty regarding the initial conditions of the system and the tidal and BYORP parameters, but we explore it in Section~\ref{sec:timescales}. 

As has been mentioned before, a positive BYORP coefficient leads to growth of the mutual orbit. The rate of the growth of the mutual semi-major axis is again the linear addition of the tidal evolution~\citep{Goldreich:2009ii} and the BYORP effect evolution~\citep{McMahon:2010jy}
\begin{equation}
\dot{a} = \frac{k_p}{Q} \left( A_T +  a^7 A_B  \right) a^{-11/2} 
\label{eqn:semimajoraxis}
\end{equation}
where the ratio of tidal parameter  $k_p/Q$ has been extracted, so that the coefficient $A_B$ depends on the BYORP coefficient $B_s$ and the tidal parameters in the exact relationship found in the tidal-BYORP equilibrium $B_s Q / k_p = 2557\ \left( R_p /1 \text{ km} \right)$ and $A_T$ depends only on better measured system properties~\citep{Jacobson:2011hp}. This ratio $k_p/Q$ is the denominator of the unknown tidal ratio $X$ described later, and is the primary source of uncertainty for both semi-major axis and eccentricity evolution equations. Since both time derivatives depend linearly on $k_p/Q$, it is conceptually easy to understand how changes in this ratio effect the evolution rates and their associated timescales.

\subsection{Adiabatic Invariance}
Using a model of a larger sphere for the primary and a smaller triaxial ellipsoid for the secondary, which is consistent with observations~\citep{Pravec:2007fw}, we derive in Appendix~\ref{sec:derivationoftheadiabaticinvariance} an adiabatic invariant relationship from the Hamiltonian dynamics between the libration amplitude and the semi-major axis (or mean motion) of the system. When a Hamiltonian dynamical system has been canonically transformed to action-angle coordinates, an adiabatic invariance is the conservation of the action (or phase space volume) of the system. The invariance is conserved even when the parameters describing the system undergo changes as long as those changes are slow compared to the periodic motion of the angle. The outward expansion of the system and the damping of the libration angle are both slow compared to the libration frequency (i.e the angle). 

The adiabatic invariance for this system is:
\begin{equation}
J_{\phi_s} = \frac{4   \mathsf{C} I_s \omega_d \mathsf{G}  }{a^{3/2}} \sqrt{\frac{3 S}{1+s}}  
\end{equation}
where $\mathsf{G} = \mathsf{G}(\sin^2 \Phi_s)$ is a function of complete elliptic integrals as given in Appendix~\ref{sec:derivationoftheadiabaticinvariance} with $\sin^2 \Phi_s$ always as the argument, $\Phi_s$ is the libration amplitude, $\mathsf{C} I_s$ is the maximum moment of inertia  of the secondary, $S$ is a shape factor of the secondary where $S=0$ is an oblate body and $S$ increases with prolateness towards $1$, $s = \mathsf{C} \mathcal{I} / \nu q a^2$ is the secondary perturbation term, and $\mathcal{I} = I_s / I_p$ is the ratio of the secondary $I_s = M_s R_s^2$ and primary moment of inertia factors  $I_p = M_p R_p^2$. The secondary perturbation term grows with increasing mass ratio $q$ or when the moment of inertia of the secondary increases relative to the primary, and it also increases with decreasing semi-major axis, however $s$ is always nearly zero; please examine Appendix~\ref{sec:derivationoftheadiabaticinvariance} for further elaboration regarding all of these terms and the derivation itself.

An intuitive understanding of the libration amplitude growth can be developed by studying the adiabatic invariant of the system at two different times:
\begin{equation}
\frac{\mathsf{G}(\sin^2 \Phi_1)}{\mathsf{G}(\sin^2 \Phi_2)}  =  \frac{a_1^{3/2} \left(1+s_1\right)^{1/2} }{a_2^{3/2}\left(1+s_2 \right)^{1/2} }
\end{equation}
where the subscripts 1 and 2 beneath the libration angle amplitude $\Phi$ and the semi-major axis $a$ indicate the two different times. This relationship can be simplified further by recognizing that $\mathsf{G}(\sin^2 \Phi_s)$ is well approximated by $\sin^2 \Phi_s$ and that $s \sim 0$ regardless of semi-major axis. Then the relationship is simply:
\begin{equation}
 \frac{\sin^2 \Phi_1}{\sin^2 \Phi_2} \propto \frac{a_1^{3/2}}{a_2^{3/2}} =  \frac{n_2}{n_1}
\end{equation}
From the relationships above, it is now clear that as the system expands, the libration amplitude increases as well.

\begin{figure}[tb!]
\begin{center}
\includegraphics[width=\columnwidth]{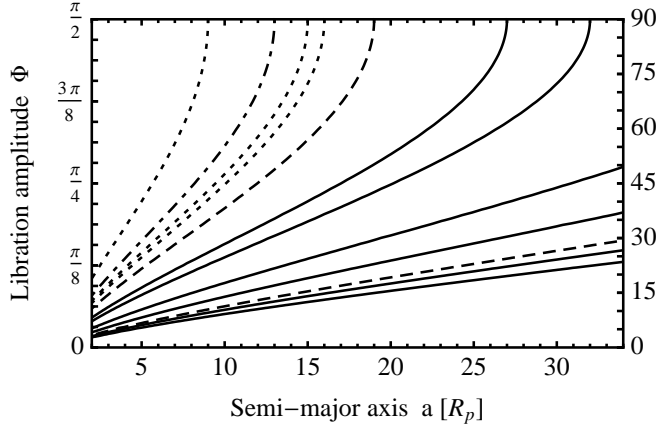}
\end{center}
\caption{Contours are $a_\text{circ} = a_\text{obs}$ for each wide asynchronous candidate as a function of $a$ and $\Phi$ using Equation~\ref{eqn:aonset}. From top/left to bottom/right the systems are 51356 (2000 RY$_{76}$), 153591 (2001 SN$_{263}$), 1717 Arlon, 1998 ST$_{27}$, 136617 (1999 CC), 317 Roxane, 32039 (2000 JO$_{23}$), 1509 Esclangona, 22899 (1999 TO$_{14}$), 3749 Balam, 17246 (2000 GL$_{74}$) and  4674 Pauling. The lines are coded such that the three tightest wide binary systems (51356, Arlon and 1998 ST$_{27}$) are dotted lines, two triples systems are dashed (Balam and 136617) while 153591 is dot-dashed since it is a tight triple system, and the rest are solid black lines.}
\label{fig:onsetcontourplots}
\end{figure}
If the libration amplitude of the secondary exceeds $90^\circ$, then the secondary will no longer be librating and it will begin to circulate. This will turn off the BYORP effect since the secondary will no longer be in a synchronous orbit. Since the tides on the primary are very weak at large semi-major axes, this effectively ends outward expansion (details are shown in Section~\ref{sec:individualsystemevolution} and particularly Figure~\ref{fig:widetidesplot}). The mutual orbit is then stranded at this semi-major axis since tides are so weak that the secondary cannot be expected to be re-locked and without the spin-orbit resonance the BYORP effect averages to zero. The YORP effect controls the rotation evolution of the asteroid and this is discussed in Section~\ref{sec:timescales}, but the mutual orbit evolution is complete. Therefore, the semi-major axis, at which the secondary begins circulating, $a_\text{circ}$ is the same semi-major axis that observers measure $a_\text{obs}$, which are the semi-major axes in Tables~\ref{tab:binaryparameters} and~\ref{tab:tripleparameters}.

To understand this evolution, we assume a given libration amplitude $\Phi$ at some semi-major axis $a$, then the semi-major axis $a_\text{circ}$ at which the secondary begins to circulate (i.e. $\Phi_\text{circ} = \pi / 2$) is:
\begin{equation}
a_\text{circ} = a \sin^{-\frac{4}{3}} \Phi
\label{eqn:aonset}
\end{equation}
In Figure~\ref{fig:onsetcontourplots}, we show contours of $a_\text{circ}$ at semi-major axes consistent with the range of observed wide asynchronous binaries. This analysis indicates that this is a plausible mechanism for creating the wide asynchronous population since all the observed singly synchronous binaries correspond to systems with small libration angles at semi-major axes consistent with formation from rotational fission. 

We have found an intuitive understanding of the adiabatic invariant relationship and the values reported above for $a_\text{init}$ and $\Phi_{init}$ are informative, but we are left wondering, `What are the real values for the initial semi-major axis and libration amplitude for each system? Do they match these predictions?' 

The semi-major axis of the system at synchronization seems a natural choice for the initial semi-major axis. As discussed previously,~\citet{Jacobson:2011eq} found that rotational fission forms stable binaries from $2$ $R_p$ to $17$ $R_p$ and possibly up to $34$ $R_p$. Tidal synchronization of the secondary is much faster than any tidal evolution of the orbit~\citep{Goldreich:2009ii,Fang:2012fw}, so this range of semi-major axes seems appropriate as possible initial values. We note that numerical simulations only produced binaries in the subset between $2$ and $17$ $R_p$.

The initial libration amplitude is more difficult to ascertain. In fact, the model so far incorrectly asserts that the libration angle should start small. As the secondary is tidally locked, it does not immediately have a small free libration, instead it has a libration amplitude near $90^\circ$. Tides from the primary acting on the free libration of the secondary dissipate energy and drive the free libration towards zero.  In the model presented so far, the free libration grows due to the adiabatic invariance through tidally and BYORP effect driven orbit expansion and it grows without any interference. The model must be expanded to include the libration dissipation due to these tides on the secondary.

\subsection{Tidal libration damping}
We directly determine the libration amplitude evolution from the energy dissipation due to the primary raising tides on the free librations of the secondary by utilizing a formulation developed by~\citet{Wisdom:2004dy,Wisdom:2008gr}. The derivation with some explanation is included in Appendix~\ref{sec:derivationoftheenergydissipationrateduetolibrationtides}. It uses a two sphere model assuming a circular orbit and homogenous interiors. It is important to note, that tides on the the free libration do not secularly evolve the orbit. This is because the tidal bulge on the secondary oscillates from the leading to trailing hemispheres and back, so the angular momentum transfer to the orbit averages to zero. 

From Appendix~\ref{sec:derivationoftheenergydissipationrateduetolibrationtides}, the tidal energy dissipation rate is:
\begin{equation}
\dot{E}  =  - \frac{2 \pi k_s \rho \omega_d^2 \omega_l \Phi_s^2 R_p^5 q^{5/3}}{ Q_l a^6}
\end{equation}
where $\Phi_s$ is the libration amplitude of the secondary and $\omega_l$ is the libration frequency. The libration frequency is derived in Appendix~\ref{sec:derivationofthelibrationfrequency} and is:
\begin{equation}
\omega_l =  \frac{ \pi \omega_d }{  2 \mathsf{K} a^{3/2} } \sqrt{ 3 S \left( 1 + s \right)} 
\end{equation}
For low mass ratio systems $q < 0.2$ and small libration amplitudes $\Phi_s \ll 1$, so the libration frequency is just proportional to the mean motion: $\omega_l \approx \pi n \sqrt{3 S / 4 }$.

In Appendix~\ref{sec:derivationofthelibrationdampingduetotides}, we use this rate of energy dissipation to calculate the time derivative of the libration amplitude due to tides:
\begin{equation}
\dot{\Phi}_s = - \frac{ A_L \Phi_s^2}{  a^{9/2} \mathsf{K} \sin 2 \Phi_s} \left( \frac{k_s}{Q_l} \right)
\end{equation}
where $A_L$ is a useful organizational coefficient reported in the Appendix. Note that the tidal parameters $k_s$ and $Q_l$ are the primary sources of uncertainty and we have left them outside of the coefficient $A_L$ on purpose. They are the numerator of the soon to be introduced tidal ratio $X$.

\subsection{Libration amplitude evolution}
By taking the time derivative of the adiabatic invariant, we determine the libration growth due to orbit expansion:
\begin{equation}
\dot{\Phi}_s  = \frac{\dot{a} \mathsf{G} A_A  }{a \mathsf{K} \sin 2 \Phi_s}
\label{eq:growth}
\end{equation}
where we have defined $A_A$ as a convenient coefficient always nearly equal to $3$ and $\mathsf{K} = \mathsf{K}(\sin^2 \Phi_s)$ is the complete elliptic function of the first kind. A derivation is found in Appendix~\ref{sec:Derivationofthelibrationgrowthduetoorbitexpansion}. 

The libration amplitude evolution has two components, the tidal damping derived above and the growth term from the adiabatic invariance. We add the two components together to get the combined effect:
\begin{equation}
\dot{\Phi}_s  = \frac{\dot{a} a \Phi_s^2}{\mathsf{K} \sin 2 \Phi_s}  \left[ \frac{A_A \mathsf{G}}{a^2 \Phi_s^2} -   \frac{ k_s  A_L  }{Q_l   \dot{a} a^{11/2}}  \right]
\label{eqn:librationamplitudetime}
\end{equation}
where $\dot{a}$ is the orbit expansion rate due to tides and the BYORP effect. 

From the formulation above, we change variables switching from the time derivative of the libration amplitude to the derivative of the libration amplitude as a function of semi-major axis:
\begin{equation}
\frac{d\Phi_s}{da} = \frac{ a \Phi_s^2}{\mathsf{K} \sin 2 \Phi_s}  \left[ \frac{ A_A \mathsf{G}}{a^2 \Phi_s^2} -  \frac{X A_L }{ A_T + a^7 A_B } \right]
\label{eqn:bigquestion}
\end{equation}
where each $A$ is a coefficient\footnote{These coefficients are independent of the semi-major axis $a$, if we assume that the secondary perturbation term $s \sim 0$. We do not make this assumption for the numerical calculations displayed in the figures, but throughout we notify the reader when it's useful to do so to better conceptually understand the math.}, $\mathsf{G}$ and $\mathsf{K}$ are functions of complete elliptic functions with $\sin^2 \Phi_s$ as the argument, and:
\begin{equation}
X = \frac{\left(  k_s / Q_l \right)}{\left( k_p / Q \right)} 
\end{equation}
which is a ratio of the tidal coefficients describing the mutual orbit evolution due to tides raised by the secondary on the primary and the tidal coefficients describing the libration evolution due to tides raised by the primary on the secondary. $X$ is the ratio of the normalized strengths of each tide normalized such that absolute scales such as size and separation found in the tidal equations are not considered. Note that the tidal Love and quality numbers are not necessarily size and frequency independent. There is much ongoing research into the determination of these values~\citep[e.g.][]{Greenberg:2009kt,Goldreich:2009ii,Jacobson:2011hp,Efroimsky:2012je,FerrazMello:2013dq}. 

Not only are the tidal parameters $k$ and $Q$ likely dependent on the physical properties of each body, but the tides on each body are fundamentally different. The tidal bulge raised by the secondary on the rapidly rotating primary moves through the body at a constant rate and direction characterized by the difference between the rotation frequency of the primary and the mean motion if we assume a circular orbit and relaxed spin state. On a spherical and homogenous primary, the shape and amplitude of the bulge stays fixed while each element of the body is distorted as it rotates through the different parts of the tidal potential. However, the tidal bulge raised by the primary on the librating secondary changes shape and amplitude as the secondary's rotation rate relative to its mean motion slows and speeds up according to its librating motion. While we have characterized these tides using similarly named parameters, the tides themselves are fundamentally different and it will require significant future geophysical and tidal modeling to assess estimates of these parameters. We can use observations to make a prediction regarding the value of $X$ assuming that most of the wide asynchronous binaries underwent this expanding singly synchronous mechanism.

\section{Comparison to observation}
\label{sec:comparisontoobservation}
We now have a model that includes the complete libration amplitude evolution, and incredibly it has only a single unknown parameter $X$. Observations provide each of the unknowns coefficients necessary to establish $A_T$, $A_B$, $A_A$ and $A_L$ with the exception of the shape factor of the secondary $S$. Since none of the secondaries of the wide asynchronous systems have published shape models, $S$ cannot be calculated for any of these systems. Instead we use the shape information corresponding to the secondary of 1999 KW$_4$, one of the best studied singly synchronous binary systems. It has a shape factor $S = 0.207$~\citep{Ostro:2006dq}. For reference, an oblate asteroid has a shape factor $S = 0$ and 1620 Geographos, the most elongated asteroid yet observed, is $S = 0.768$~\citep{Ostro:1996io}. Since 1999 KW$_4$ is a singly synchronous binary, its shape factor is the most appropriate an an initial guess. The final results are insensitive to a factor of two or three uncertainty in the value of $S$ since it appears only as a square root in $A_L$.

\begin{figure}[tb]
\begin{center}
\includegraphics[width=\columnwidth]{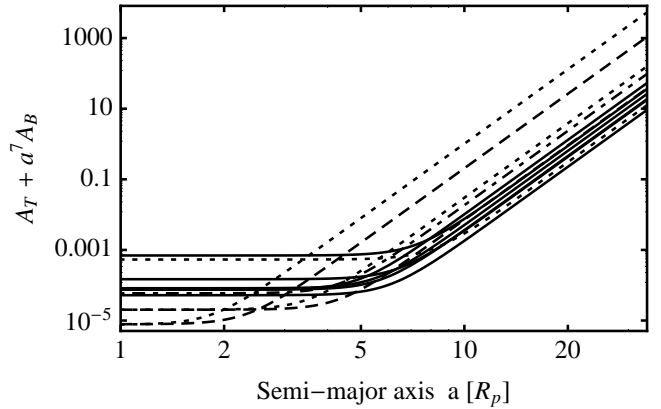}
\end{center}
\caption{The term $A_T + a^7 A_B$ plotted as a function for all of the wide asynchronous systems. The lines are from top to bottom along the right hand side are 1998 ST$_{27}$, 136617 (1994 CC), 51356 (2000 RY$_{76}$), 153591 (2001 SN$_{263}$), 4674 Pauling, 3749 Balam, 32039 (2000 JO$_{23}$), 1509 Esclangona, 22899 (1999 TO$_{14}$), 17246 (2000 GL$_{74}$), 1717 Arlon and 317 Roxane. The lines are coded such that the three tightest wide binary systems (51356, Arlon and 1998 ST$_{27}$) are dotted lines, two triples systems are dashed (Balam and 136617) while 153591 is dot-dashed since it is a tight triple system, and the rest are solid black lines.}
\label{fig:Abplot}
\end{figure}
Another observation provided parameter is the tidal-BYORP effect equilibrium parameter relationship $B_s Q / k_p = 2557\ \left( R_p /1 \text{ km} \right)$~\citep{Jacobson:2011hp}. This is the greatest source of uncertainty in the model outside of the parameter $X$ since $B_s Q / k_p$ has more than an order of magnitude uncertainty. We can study the influence of this uncertainty by noticing that $B_s Q / k_p$ appears as a linear factor in $A_B$ and examining the role of $A_B$ in Equation~\ref{eqn:bigquestion} in detail. In Figure~\ref{fig:Abplot}, we show the denominator of the second term in the brackets of Equation~\ref{eqn:bigquestion}: $A_T + a^7 A_B$ as a function of semi-major axis. It is clear that for semi-major axes $a \gtrsim 8 $ $R_p$ the term $a^7 A_B$ dominates and the $A_T$ term can be neglected. This is exactly what has been argued all along, tides are extraordinarily weak compared to the BYORP effect at large semi-major axes. Naturally, the turning point in Figure~\ref{fig:Abplot} for each of these systems is the semi-major axis equilibrium that the tidal-BYORP equilibrium theory predicts would exist if these systems were BYORP effect contractive rather than expansive. Encouragingly, all the turning points are located interior of the observed orbits of the wide asynchronous candidates. Also these points are in the same region of semi-major axis space that the observed singly synchronous binaries occupy with the exception of 1998 ST$_{27}$, whose predicted equilibrium is a very tight binary system $a \lesssim 2$ $R_p$. Adjusting $B_s Q / k_p$ moves the turning points. Decreasing/increasing $B_s Q / k_p$ moves the turning points outward/inward. Although even a change by an order of magnitude does not move them very far since the equilibrium goes as $(A_T/A_B)^{1/7}$.

When the binary system is tight (i.e. below the turning point) the evolution of the libration amplitude is insensitive to the choice of $B_s Q / k_p$. This can be understood as a regime dominated by the two sets of tides. The libration tides on the secondary damping the libration amplitude and the tidal expansion due to tides on the primary growing the amplitude through the adiabatic invariance.

When the wide asynchronous binaries are evolving exterior to the turning points, $A_T$ is ignorable. Then Equation~\ref{eqn:bigquestion} becomes simply $d \Phi_s / d a \propto X / \left( B_s Q / k_p \right)$ when all other parameters are held constant. As we explore how the libration amplitude, the initial semi-major axis and the circulation semi-major axis vary as a function of X, we can keep this relationship in mind to better understand how the uncertainty in $B_s Q / k_p$ effects the results.  

\subsection{Individual system evolution}
\label{sec:individualsystemevolution}
\begin{figure}[tb]
\begin{center}
\includegraphics[width=\columnwidth]{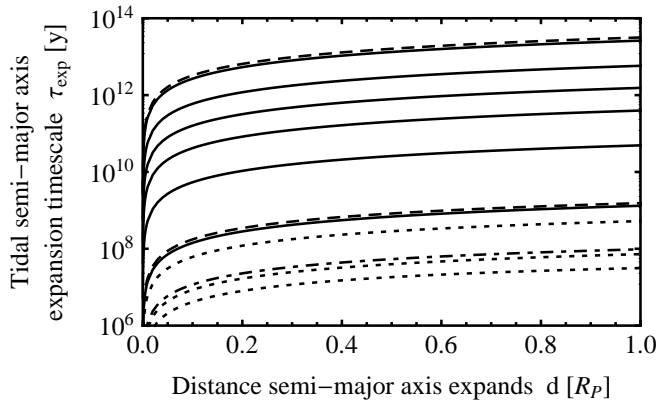}
\end{center}
\caption{The time necessary for tides to expand the orbit of each wide asynchronous binary to its current observed semi-major axis from some distance interior of that orbit indicated along the abscissa. The timescale is calculated by numerically integrating: $ \tau_\text{exp} = \int_{a_\text{obs}-d}^{a_\text{obs}} \left( \frac{Q}{k_p} \right) \frac{a^{11/2}}{A_T}\ da $
where the role of the tidal coefficients $Q$ and $k_p$ is highlighted since they are the major source of uncertainty. $Q/k_p$ is estimated from the tidal-BYORP effect equilibrium assuming a BYORP coefficient of $B_s = 2 \times 10^{-2}$. The lines in order from top to bottom are 3749 Balam, 4674 Pauling, 17246 (2000 GL$_{74}$), 22899 (1999 TO$_{14}$), 1509 Esclangona, 317 Roxane, 136617 (1994 CC), 32309 (2000 JO$_{23}$), 1998 ST$_{27}$, 153591 (2001 SN$_{263}$), 1717 Arlon and 51356 (2000 RY$_{76}$). The lines are coded such that the three tightest wide binary systems (51356, Arlon and 1998 ST$_{27}$) are dotted lines, two triples systems are dashed (Balam and 136617) while 153591 is dot-dashed since it is a tight triple system, and the rest are solid black lines.}
\label{fig:widetidesplot}
\end{figure}
As has been explained in detail earlier in this work, when the libration of the secondary reaches $90^\circ$ it begins to circulate. This circulation turns off the BYORP effect. Mutual body tides continue to act on the system, but these tides are very weak. Figure~\ref{fig:widetidesplot} shows the time it takes for each wide asynchronous candidate system to expand to its current semi-major axis from a semi-major axis smaller by a distance $d$. Even for the tightest of these systems the time necessary to expand a single primary radius is $\sim100$ Myr. It is clear that if these systems expanded after circulation of the secondary turned off the BYORP effect, they did not expand very much. These systems are stranded and the semi-major axis the binary is observed at is very close to the semi-major axis at which the secondary began to circulate. 

\begin{figure*}[tb]
\begin{center}
\includegraphics[width=\textwidth]{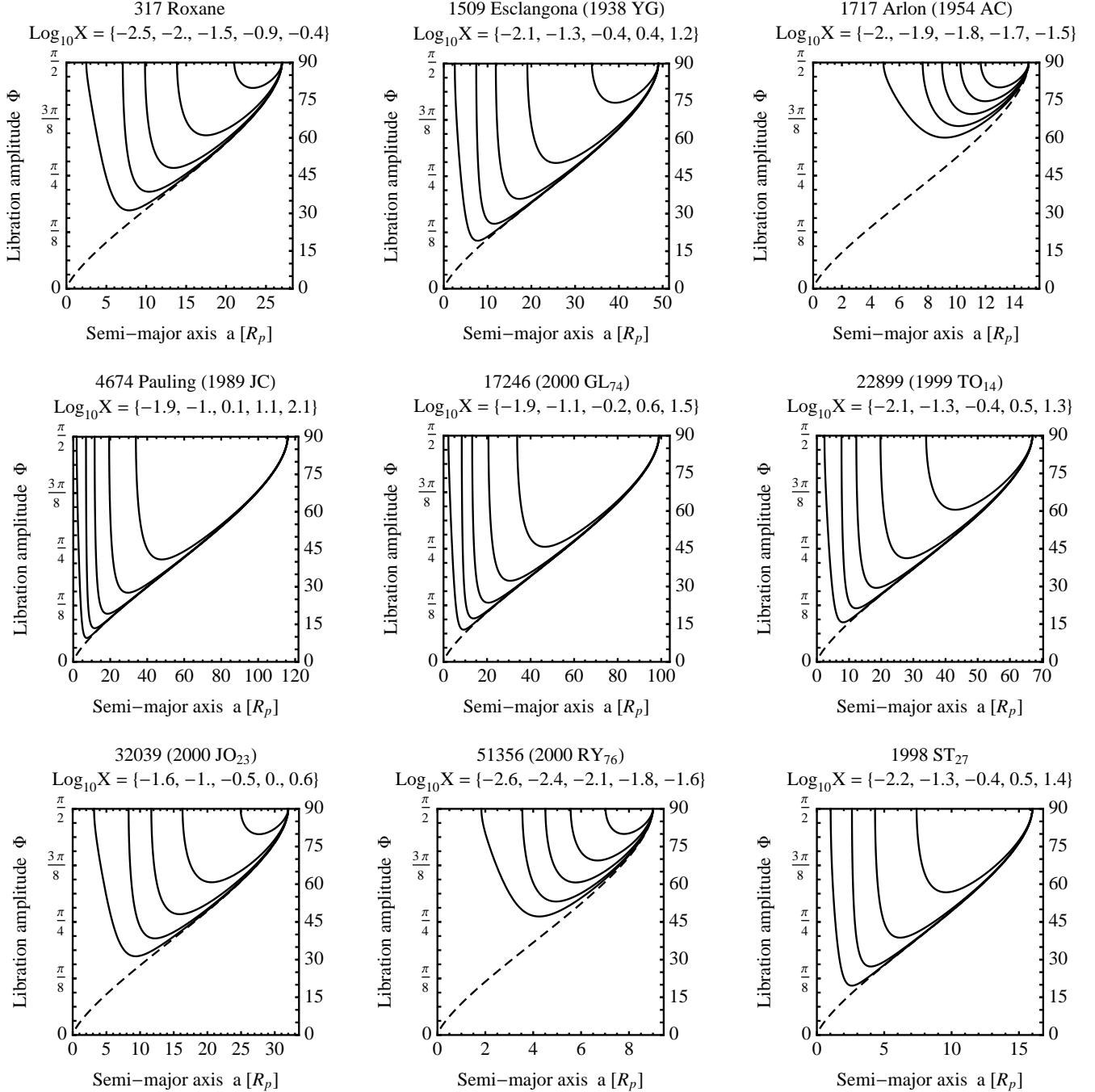}
\end{center}
\caption{Possible evolutionary tracks for each of the nine candidate wide asynchronous binary systems. Each plot shows the libration amplitude as a function of semi-major axis for a specific system. The different lines represent a range of tidal ratios $X$ listed above each frame in the same order as the lines left-to-right. The tidal ratios are chosen for each system to demonstrate the range of initial semi-major axes that can lead to the observed semi-major axis of the system. The initial semi-major axes are constrained to be between $2$ and $30$ $R_p$, which are the limits from~\citet{Jacobson:2011eq}. The dashed line is an asymptote explained in the text.}
\label{fig:individualplots}
\end{figure*}
Assuming a value of $X$, we numerically integrate Equation~\ref{eqn:bigquestion} to the observed semi-major axis of that system to uncover the libration amplitude history and the initial synchronization semi-major axis. The results of these integrations are shown for all nine candidate wide asynchronous binaries in Figure~\ref{fig:individualplots}. We have chosen to show a selection of five possible evolutionary tracks for each system. Each track has a different tidal ratio $X$ and each is listed above the plot under the name of the system. They correspond to the lines from left to right. Under different assumptions of $X$ each system can form as a stable binary and tidally synchronize its secondary within the range of semi-major axes ($2 < a < 34$ $R_p$) consistent with the observed singly synchronous and tight asynchronous populations and the simulated stable binaries from~\citet{Jacobson:2011eq}. Then the binary expands from those separations to their observed semi-major axes. During that expansion, the secondary remains tidally locked but freely librating. 

A libration amplitude evolution pattern emerges across each of these systems. At the tidal synchronization semi-major axis, the libration amplitude is $90^\circ$. The libration amplitude declines due to tides as the system expands only a small amount. The slope of the decline is controlled by the second term in Equation~\ref{eqn:bigquestion}: $-X A_L / \left( A_T + a^7 A_B\right)$. The shallower slopes at small semi-major axes are due to the relative similarity of the $A_T$ and $a^7 A_B$ terms at those small semi-major axes. For systems with larger tidal synchronization semi-major axes, the slope is much steeper due to the dominance of the $a^7 A_B$ term. Shallow or steep, the libration amplitude damps until the first and second terms of Equation~\ref{eqn:bigquestion}  balance, adiabatic growth and tidal damping.

Equation~\ref{eqn:bigquestion} is a highly nonlinear equation, so it is only possible to exactly solve for this equilibrium point numerically. However we can approximately solve for this equilibrium semi-major axis $a_{eq}$, by making the following assumptions: (1) the secondary perturbation term $s \sim 0$ and (2) that $a$ is sufficiently large so that $A_T \ll a^7 A_B$. Then,
\begin{equation}
a_{eq} = \left( \frac{ X A_L \Phi_s^2}{A_A A_B \mathsf{G} } \right)^{1/5}
\label{eqn:equilibrium}
\end{equation}
At this semi-major axis each of the curves in Figure~\ref{fig:Abplot} turn around, approach an asymptote, then follow that asymptote to the observed semi-major axis of the system. This asymptote is Equation~\ref{eq:growth} integrated. Using the same approximation that the secondary perturbation term $s \sim 0 $, then the asymptote is defined by
\begin{equation}
a = a_\text{circ} \exp \left[ - \int_\phi^{\pi/2} \frac{\mathsf{K} \sin 2 \Phi_s}{3 \mathsf{G}}\ d\Phi_s \right]
\end{equation}
If it is discovered from future observations that a wide binary is librating, then this asymptote may be used to assess the future evolution of the system. The previous equilibrium equation determines whether it is approaching the asymptote or the equilibrium. This could provide a significantly better understanding of the tidal ratio $X$ than we now possess. Studying such systems would be very useful. 185851 (2000 DP$_{107}$) is the widest tight synchronous binary and an outlier in the tidal-BYORP equilibrium dataset and so therefore a candidate singly synchronous expanding system with a librating secondary~\citep{Jacobson:2011hp}.

\begin{figure*}[tb!]
\begin{center}
\includegraphics[width=\textwidth]{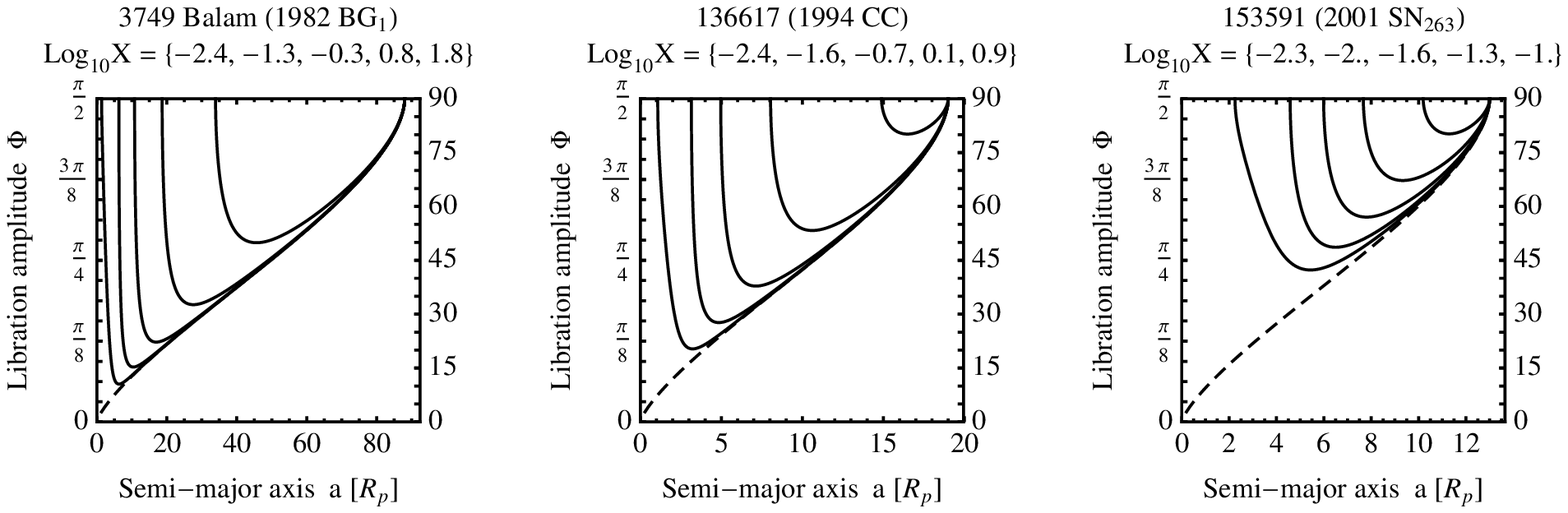}
\end{center}
\caption{Identical to Figure~\ref{fig:individualplots} but for the candidate wide asynchronous triple systems.}
\label{fig:individualtripleplots}
\end{figure*}
Repeating this analysis with the outer pairs of the triple systems, we find similar results. These are shown in Figure~\ref{fig:individualtripleplots}. Each pair could have formed in the $2$ to $34$ $R_p$ region then expanded to its current orbit. 

\begin{figure*}[tb!]
\begin{center}
\includegraphics[width=\textwidth]{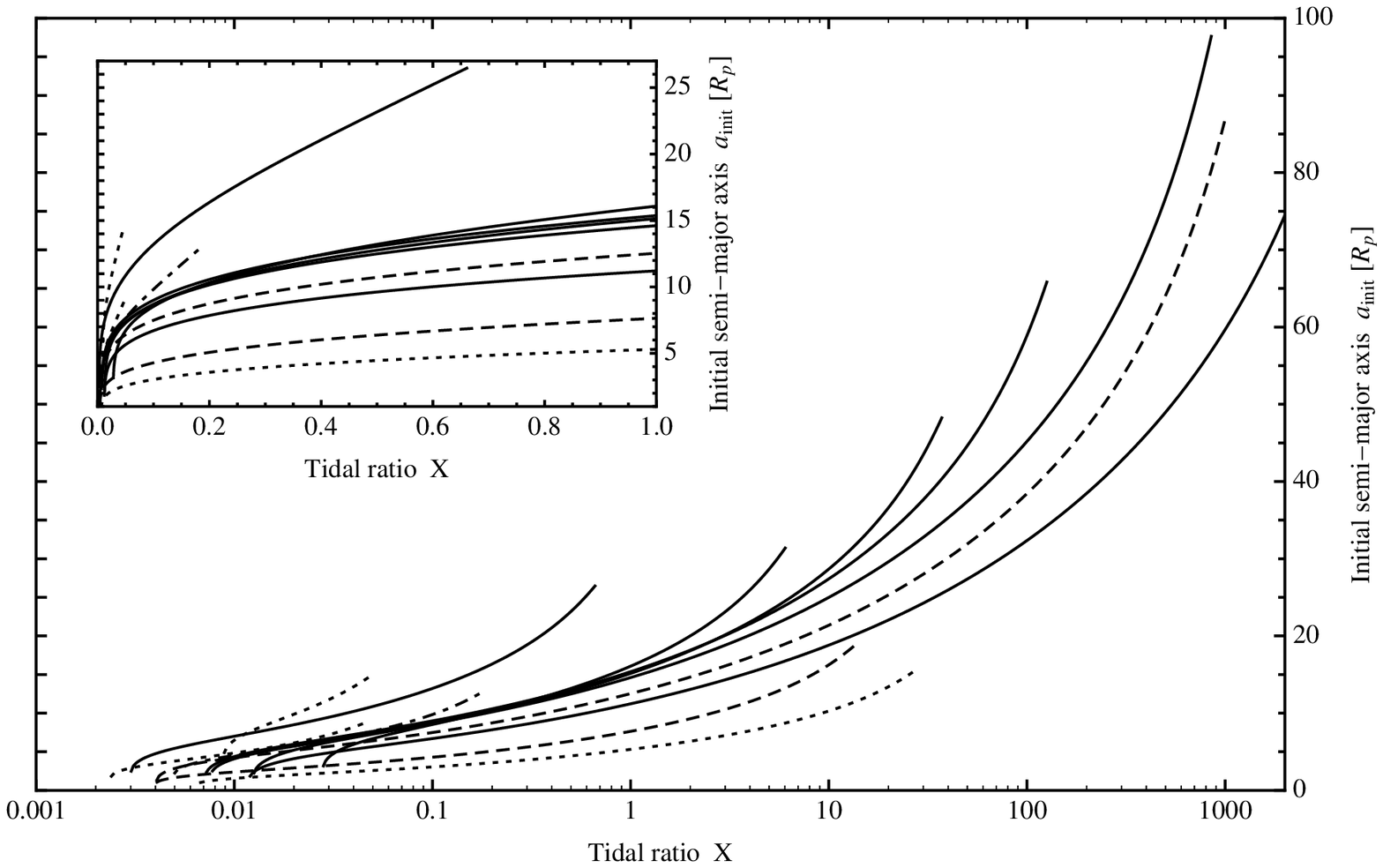}
\end{center}
\caption{The initial or synchronization semi-major axis for each binary system as a function of the tidal ratio $X$. The inset figure is the same, but with a linear tidal ratio $X$ axis and focused on the region where normalized libration tides are weaker than normalized circulation tides $X < 1$. From the left/top to right/bottom the lines are 1717 Arlon, 317 Roxane, 51356 (2000 RY$_{76}$), 153591 (2001 SN$_{263}$), 32039 (2000 JO$_{23}$), 1509 Esclangona, 22899 (1999 TO$_{14}$), 17246 (2000 GL$_{74}$), 3749 Balam, 4674 Pauling, 136617 (1994 CC) and 1998 ST$_{27}$. The lines are coded such that the three tightest wide binary systems (51356, Arlon and 1998 ST$_{27}$) are dotted lines, two triples systems are dashed (Balam and 136617) while 153591 is dot-dashed since it is a tight triple system, and the rest are solid black lines.}
\label{fig:xplot}
\end{figure*}
In each of the Figures~\ref{fig:individualplots} and~\ref{fig:individualtripleplots}, we have only shown evolutionary paths corresponding to synchronization semi-major axes between $2$ and $34$ $R_p$. For systems with observed semi-major axes (i.e. circulation semi-major axes) $a_\text{circ} > 34$ $R_p$, there exist solutions all the way from $34$ $R_p$ to $a_\text{circ}$ with different values of $X$. Figure~\ref{fig:xplot} shows how the initial semi-major axis for each system depends on the tidal ratio $X$. There is a one-to-one relationship between the tidal ratio $X$ and the synchronization (initial) semi-major axis for each system; as the tidal ratio $X$ increases, the initial semi-major axis increases. 

From Figure~\ref{fig:xplot}, some intuition about the tidal ratio $X$ can be gleaned if we assume that the wide asynchronous binaries formed from this expanding singly synchronous mechanism. Before we begin to speculate, it is important to remember that it is very unlikely that the tidal ratio $X$ should have the same value across all systems. Likely it will depend on the absolute sizes of the binary members, their compositions and the relevant frequencies: mean motion $n$, primary spin rate $\omega_p$ and secondary libration frequency $\omega_l$. Since these frequencies change, assuming a constant tidal ratio across the entire evolutionary history is a zeroth order assumption likely to be violated, however we defend it as an approximation by noting that the equilibrium semi-major axis defined in Equation~\ref{eqn:equilibrium} is only weakly dependent on the tidal ratio $a_{eq} \propto X^{1/5}$, and the post-equilibrium asymptote is not dependent on the tidal ratio. 

Figure~\ref{fig:xplot} shows that if $X \sim 0.1$ then $10$ out of $12$ systems start within the $2$ to $17$ $R_p$ range and expand to their observed semi-major axes. The two that do not are both tight systems (51356 (2000 RY$_{76}$) and 1717 Arlon). And in general, if $0.01 < X < 1$, then most systems start within the appropriate range and expand to their circulation semi-major axes. Since $X$ is the ratio of the normalized strength of the libration tide to the circulation tide, then we might expect the libration tide to be weaker because it is not a constant tide moving through a body like the circulation tide. As described before, the libration tide waxes and wanes as the rotation rate of the secondary speeds up and slows down relative to the mean motion, and so $X < 1$ may be a naturally expected outcome.

It is unclear what the fraction of expanding singly synchronous systems that desynchronize is relative to the total number that expand out towards the Hill radius. Those asteroids that do not desynchronize before the Hill radius would be considered asteroid pairs after solar perturbations disrupt the mutual orbit. These are not the asteroid pairs commonly found~\citep{Pravec:2010kt}. Those pairs are created directly from the chaotic and inefficient process of forming a stable binary after a rotational fission event.~\citet{Jacobson:2011eq} estimates that only $\sim8\%$ of systems undergoing rotational fission result in a stable binary, the rest of the systems form asteroid pairs some with more than one member. Of that small minority that form stable binaries, nominally half will have expansive BYORP torques and expand towards the Hill radius and disruption. This would create a population of asteroid pairs with one rapidly and one slowly rotating member, but only one such system for every $23$ asteroid pairs that obey the relationship in~\citep{Pravec:2010kt}. This ratio will be a difficult observation to make.

We have shown it is possible to get from a stable binary formed from a rotationally fission event to a wide asynchronous binary. The expanding singly synchronous mechanism works, but is it plausible? Part of its plausibility rests with determining the proper value of the tidal ratio $X$, which will take significant future work. Lets assume the tidal ratio $X$ takes a value that is consistent with the expanding singly synchronous mechanism, then the binary evolves to a wide asynchronous state, but does it do so on a reasonable timescale?

\subsection{Evolutionary phases and timescales}
\label{sec:timescales}
The expanding singly synchronous mechanism can be broken into three evolutionary phases. First the tidal synchronization of the secondary. Second, the BYORP effect driven expansion of the orbit. Third, the YORP effect rotational acceleration of the secondary. In order for this mechanism to be plausible, wide asynchronous binaries must evolve through all three phases within the estimated lifetimes of small binary systems, which in the near-Earth asteroid population are on the order of a few million years due to dynamical scattering into the Sun~\citep{BottkeJr:2002dp} and in the Main Belt asteroid population are on the order of a few hundred million years due to collisions amongst asteroids~\citep{BottkeJr:2005gd}. We detail each evolutionary phase in order.

\begin{figure}[tb!]
\begin{center}
\includegraphics[width=\columnwidth]{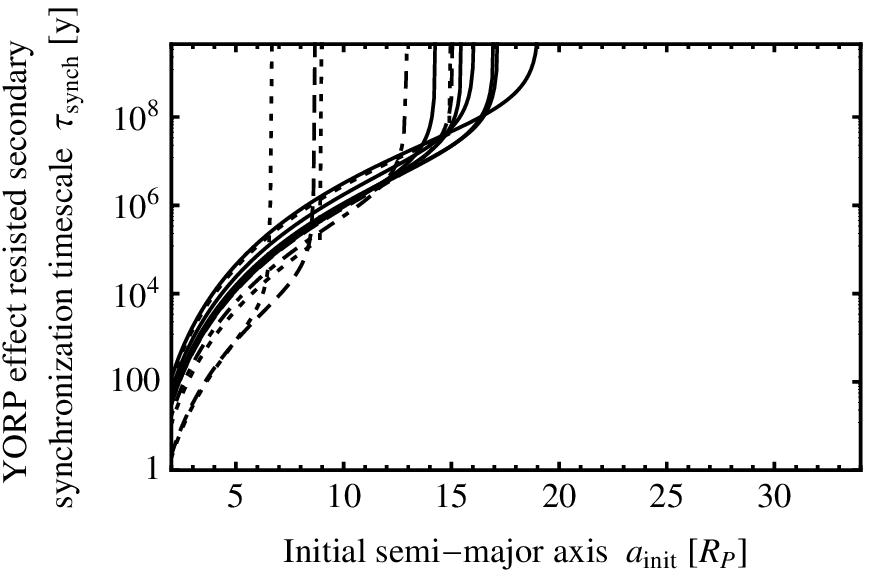} 
\includegraphics[width=\columnwidth]{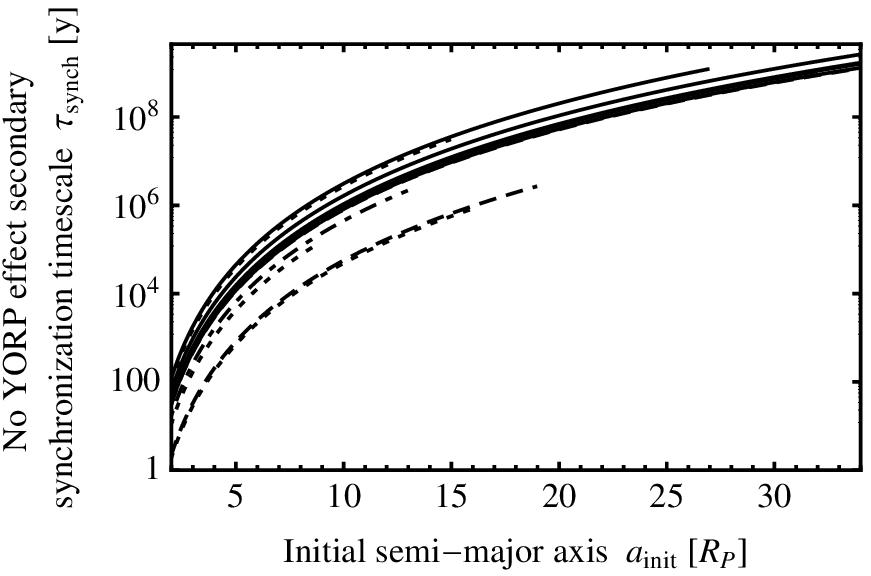}
\includegraphics[width=\columnwidth]{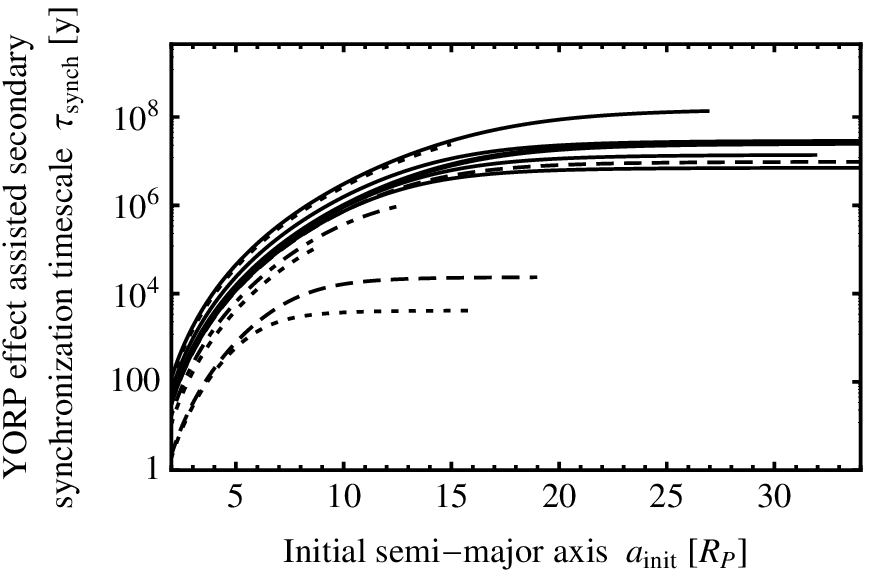}
\end{center}
\caption{These three plots show the synchronization timescale $\tau_\text{synch}$ for every wide asynchronous candidate but assuming different secondary YORP coefficients in each plot. In the top plot the YORP effect resists synchronization $Y_s = 0.02$, the middle plot there is no YORP effect $Y_s = 0$, and in the bottom plot the YORP effect assists synchronization $Y_s = - 0.02$. In the middle (no YORP effect) plot, the lines from top to bottom are 317 Roxane, 1717 Arlon, 1509 Esclangona, 17246 (2000 GL$_{74}$), 22899 (1999 TO$_{14}$), 32039 (2000 JO$_{23}$), 4674 Pauling, 3749 Balam, 153591 (2001 SN$_{263}$), 51356 (2000 RY$_{76}$), 136617 (1999 CC) and 1998 ST$_{27}$. The lines are coded such that the three tightest wide binary systems (51356, Arlon and 1998 ST$_{27}$) are dotted lines, two triples systems are dashed (Balam and 136617) while 153591 is dot-dashed since it is a tight triple system, and the rest are solid black lines.}
\label{fig:tidsynchplot}
\end{figure}
The synchronization of the secondary depends strongly on the initial semi-major axis of the binary and the YORP coefficient of the secondary as shown in Figure~\ref{fig:tidsynchplot}. The synchronization timescale is calculated by linearly adding the tidal and YORP effect torques~\citep{Goldreich:2009ii,Scheeres:2007kv}: 
\begin{equation}
\tau_\text{synch} = \left[ \frac{15 k_p \omega_d^2}{4 Q a_\text{init}^6} + \frac{Y_s H_\odot }{2 \pi \rho R_p^2 q^{2/3}} \right]^{-1} \left( \frac{\omega_{d}}{1.2} - \frac{\omega_d}{\nu^{1/2} a_\text{init}^{3/2}}  \right)
\end{equation}
where each secondary is assumed to be initially rotating near its surface disruption limit ($1.2$ $P_d$) consistent with~\citet{Jacobson:2011eq} and the YORP coefficient $Y_s$ is a dimensionless number that quantifies the asymmetric shape of the body. The YORP effect is a summation of radiative torques similar to the BYORP effect, but the lever arm of each torque is from the radiating or irradiated surface element to the body center of mass rather than the barycenter of the mutual orbit~\citep{Rubincam:2000fg}. The YORP effect can torque the secondary in the same direction (assistively) or opposite direction (resistively) as the tidal torque. These two scenarios along with the no YORP effect scenario are shown in Figure~\ref{fig:tidsynchplot}. The tidal parameters $Q/k_p$ are estimated from the tidal-BYORP effect equilibrium by assuming a BYORP coefficient of $B_s = 2 \times 10^{-2}$~\citep{McMahon:2010jy,Jacobson:2011hp}.

From this analysis, it becomes clear that there are three important regimes. First, asynchronous binaries with semi-major axes $a \lesssim 8$ $R_p$. In this regime, tides generally dominate the evolution of the secondary. Not only do tides synchronize each system but they do so quickly in typically less than a million years. In the second regime, the semi-major axes range between $\sim 8 $ and $\sim 17$ $R_p$. The YORP effect and tidal torques are nearly the same. Resistive YORP torques can prevent synchronization within this range.~\citet{Jacobson:2012tx} concluded that this resistive scenario is responsible for the observed tight asynchronous binary population. Typical timescales from synchronization are between a million and a hundred million years. Last, the widest semi-major axes $a \gtrsim 17$ $R_p$. In this range the YORP effect torques are typically much stronger than the tidal torques with the caveat that very symmetric secondaries could have nearly non-existent YORP coefficients. Wide asynchronous binaries that form directly from rotational fission exist in this range. These last two regimes can also be seen in Figure~\ref{fig:ratioplots}, where the torques are directly compared for each system's observed semi-major axis.

Given the observed wide asynchronous candidates, the analysis above, and the analysis of the tidal ratio $X$ in Section~\ref{sec:individualsystemevolution}, it seems likely that the wide asynchronous progenitors synchronize in the first and second regimes $a \lesssim 17$ $R_p$ and then de-synchronize in the second and third regimes $a \gtrsim 8$ $R_p$. If this is so, then all three near-Earth asteroid candidates synchronized in $\lesssim 10^6$ yr, and the other Main Belt candidates likely synchronized in $\lesssim 10^8$ yrs. The only exception is if the YORP effect is resistive, then for each YORP coefficient there exists a semi-major axis above which the system will never synchronize. All observed tight asynchronous binary systems exist at semi-major axes large enough to be consistent with being grater than this special semi-major axis~\citep{Jacobson:2012tx}. It's possible that systems like 1717 Arlon and 51356 (2000 RY$_{76}$) have more in common with the tight asynchronous population than the other wides asynchronous candidates. They possibly formed directly from rotational fission. Future work will focus specifically on the role of the YORP effect in binary asteroid systems bringing all of these ideas together.

\begin{figure}[tb!]
\begin{center}
\includegraphics[width=\columnwidth]{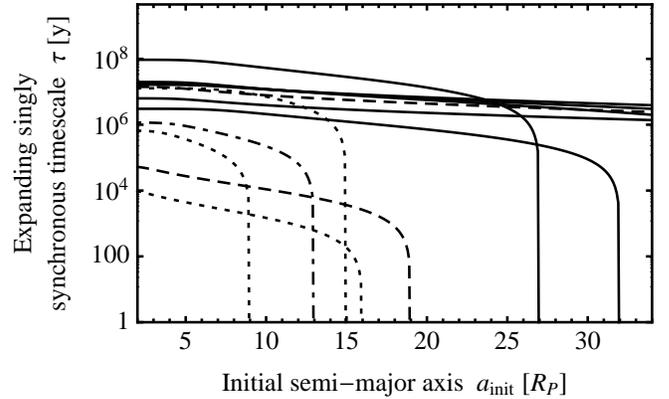}
\end{center}
\caption{The time for each binary to expand as a singly synchronous system to its observed semi-major axis is shown as a function of its initial semi-major axis. The lines from top to bottom along the lefthand side are 317 Roxane, 1509 Esclangona, 22899 (1999 TO$_{14}$), 17246 (2000 GL$_{74}$), 3749 Balam, 1717 Arlon, 4674 Pauling, 32039 (2000 JO$_{23}$), 153591 (2001 SN$_{263}$), 51356 (2000 RY$_{76}$), 136617 (1999 CC) and 1998 ST$_{27}$. The lines are coded such that the three tightest wide binary systems (51356, Arlon and 1998 ST$_{27}$) are dotted lines, two triples systems are dashed (Balam and 136617) while 153591 is dot-dashed since it is a tight triple system, and the rest are solid black lines.}
\label{fig:expsingplot}
\end{figure}
The second evolutionary phase is the orbit expansion. This is simply the numerical integration of Equation~\ref{eqn:semimajoraxis}:
\begin{equation}
\tau = \int_{a_\text{init}}^{a_\text{obs}} \left( \frac{Q}{k_p} \right) \frac{a^{11/2}}{A_T + a^7 A_B}\ da
\end{equation}
where the linear role of the tidal coefficients $Q$ and $k_p$ are highlighted since they are the major source of uncertainty. The expanding singly synchronous timescales are shown in Figure~\ref{fig:expsingplot}. All three near-Earth asteroid candidates expand in $\lesssim 10^6$ yr, and the other Main Belt candidates expand in $\lesssim 10^8$ yrs. 

The third evolutionary phase is the YORP effect rotational acceleration of the secondary. In some sense this is the complement of the first phase. However instead of driving towards or away from synchronization, the YORP effect always drives the secondary away from synchronization. In fact during the expansion phase, the YORP effect is always active on the secondary and imparts an angular offset on the libration state of the secondary, however this offset is very small with the exception of the nearly oblate secondary~\citep{Jacobson:2011vk}. This offset does determine from which side of the libration potential the secondary exits. If the YORP effect is "assistive" prior to synchronization then the YORP torque will rotationally decelerate the secondary after desynchronization. After the angular momentum of the secondary has been sufficiently drained by the YORP effect, it would be possible for small impacts to knock the body into a non-principal axis rotation state~\citep{Henych:2013hz}. None of the observed wide asynchronous binaries have secondary periods consistent with a sub-keplerian rotation rate or non-principal axis rotation, however observing such a slow (100s of hours) period or a period with multiple components is a considerable challenge. However, due to the tangential YORP effect~\citep{Golubov:2012kt}, we expect most secondaries to have a "resistive" YORP effect torque prior to synchronization. Thus, they will rotationally accelerate after desynchronization. This sequence of events is consistent with all the confirmed wide asynchronous binaries assuming that the spin and orbit poles are all aligned. Regardless of the direction of the YORP effect, mutual body tides resist the YORP effect torque after desynchronization, and the significant difference between the first and third phases is the orbit expansion of the second phase. At a much wider orbit, the tides are significantly weaker. 

 \begin{figure}[tb!]
\begin{center}
\includegraphics[width=\columnwidth]{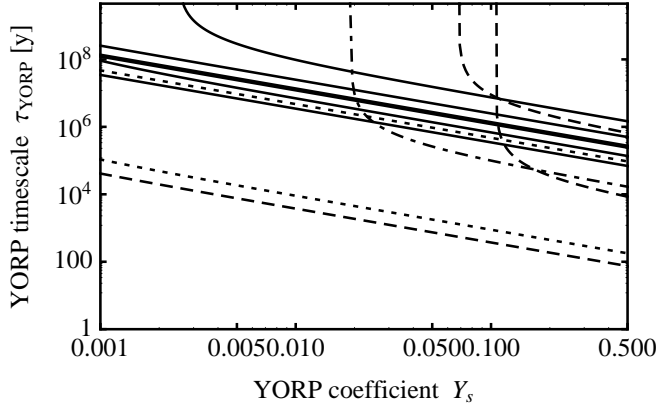}
\end{center}
\caption{The time needed for the YORP effect to rotationally accelerate the secondary to it's observed rotation rate $\omega_\text{obs}$ is shown as a function of the YORP coefficient of the secondary $Y_s$. The lines from top to bottom along the righthand side are 317 Roxane, 1717 Arlon, 1509 Esclangona, 17246 (2000 GL$_{74}$), 22899 (1999 TO$_{14}$), 32039 (2000 JO$_{23}$), 3749 Balam, 4674 Pauling, 153591 (2001 SN$_{263}$), 51356 (2000 RY$_{76}$), 136617 (1994 CC) and 1998 ST$_{27}$. The lines are coded such that the three tightest wide binary systems (51356, Arlon and 1998 ST$_{27}$) are dotted lines, two triples systems are dashed (Balam and 136617) while 153591 is dot-dashed since it is a tight triple system, and the rest are solid black lines.}
\label{fig:YORPPlot}
\end{figure}
Figure~\ref{fig:YORPPlot} shows the time necessary for the YORP effect to rotationally accelerate the secondary to its observed rotation rate $\omega_\text{obs}$ is shown as a function of the YORP coefficient of the secondary $Y_s$. Typical values are $\sim10^{-2}$~\citep{Scheeres:2007kv}. For systems without an observed rotation rate, we use the average rate of $5.5$ hrs. The timescale is:
\begin{equation}
\tau_\text{YORP} = \left[ \frac{Y_s H_\odot }{2 \pi \rho R_p^2 q^{2/3}} - \frac{15 k_p \omega_d^2}{4 Q a_\text{obs}^6}  \right]^{-1} \left( \omega_\text{obs} - \frac{\omega_d}{\nu^{1/2} a_\text{obs}^{3/2}} \right)
\end{equation}
where the tidal parameters $Q/k_p$ are estimated from the tidal-BYORP effect equilibrium by assuming a BYORP coefficient of $B_s = 2 \times 10^{-2}$~\citep{McMahon:2010jy,Jacobson:2011hp}. 

Looking at Figure~\ref{fig:YORPPlot}, the tight systems (51356 (2000 RY$_{76}$), 153591 (2001 SN$_{263}$) and 1717 Arlon) require large YORP coefficients in order to rotationally accelerate their secondaries to their observed (or in the case of 51356 (2000 RY$_{76}$), assumed) rotation rates. These systems also stood out in Figure~\ref{fig:ratioplots}. This suggests that these are systems are being tidally locked rather than accelerating away from synchronicity. Since all have semi-major axes $a < 17$ $R_p$, they are candidates for direct formation by rotational fission or in the case of triple system 153591 (2001 SN$_{263}$) something more exotic could be occurring. 317 Roxanne needs to have a non-negligible YORP coefficient in order to rotationally accelerate away from synchronicity. The other systems do not require particularly large YORP coefficients in order to be driven by the YORP effect away from synchronicity to their observed (or assumed) rotation rates in appropriate amounts of time.

\begin{figure}[tb!]
\begin{center}
\includegraphics[width=\columnwidth]{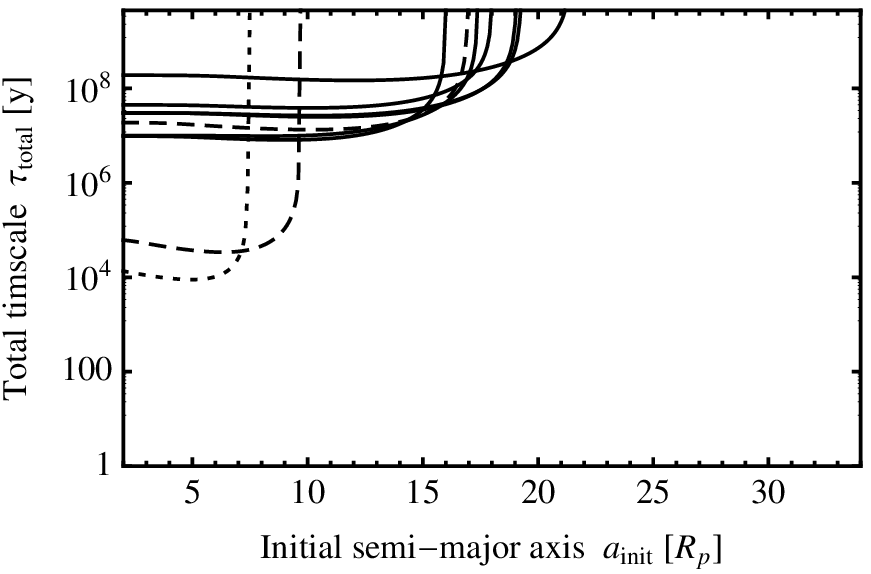}
\includegraphics[width=\columnwidth]{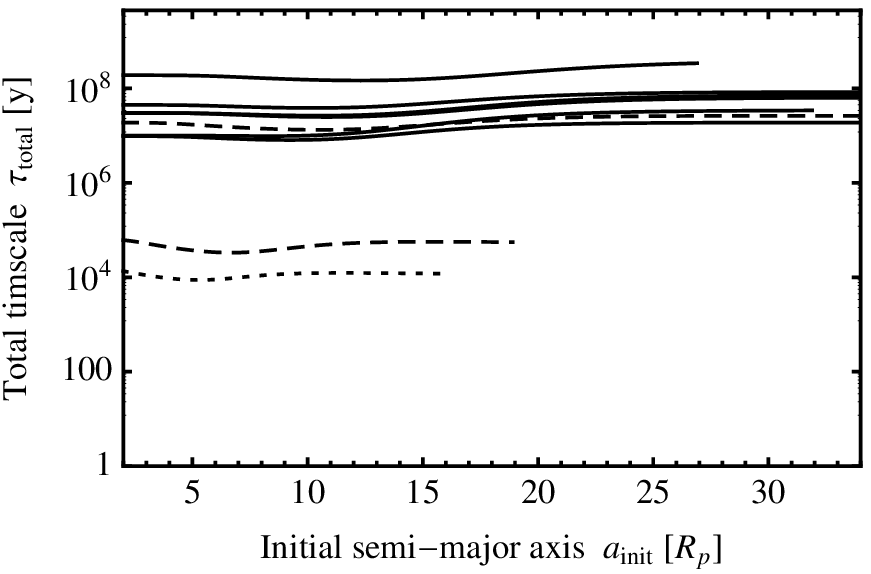}
\end{center}
\caption{The total time for each binary system to evolve through all three evolutionary phases shown in two scenarios: Above, the YORP effect on the secondary resists synchronization and below, the YORP effect on the secondary assists synchronization. The lines from top to bottom along the lefthand side of each plot are 317 Roxane, 4674 Pauling, 1509 Esclangona, 3749 Balam, 17246 (2000 GL$_{74}$), 22899 (1999 TO$_{14}$),  32039 (2000 JO$_{23}$), 136617 (1994 CC), and 1998 ST$_{27}$. The lines are coded such that 1998 ST$_{27}$ is a dotted lines, two triples systems are dashed (Balam and 136617), and the rest are solid black lines.}
\label{fig:totalplot}
\end{figure}
Lastly, we consider the combined timescales of all three phases. This is difficult to do without having measurements of the secondary YORP coefficients for each system, so we explore two scenarios: resistive and assistive, assuming a YORP coefficient for the secondary of $Y_s = 10^{-2}$ in both cases. Then, it's a simple sum of the timescales associated with each phase. The results are shown in Figure~\ref{fig:totalplot}. We do not include 51356 (2000 RY$_{76}$), 153591 (2001 SN$_{263}$) and 1717 Arlon; each of these systems is likely synchronizing and not candidates for the expanding singly synchronous mechanism.

In the total timescale results, there are two systems with very short timescales:  136617 (1994 CC) and 1998 ST$_{27}$. This makes them unlikelier candidates for the expanding singly synchronous mechanism since it implies that we are observing them at a rather special time. 317 Roxane is at the other extreme with a total timescale exceeding a $100$ Myr, however it is a larger asteroid system and its collisional lifetime is accordingly larger as well~\citep{Farinella:1998ff} so this system is still well within its expected lifetime in the Main Belt asteroid population. The six other systems also obey the lifetime constraint of their respective populations especially if they formed interior of $a \lesssim 17$ $R_P$.

\section{Discussion}
\label{sec:discussion}
Above, we have examined the theory behind the expanding singly synchronous mechanism for the formation of the wide asynchronous binaries. It is important to clarify that this mechanism is distinct from the onset of chaotic rotation of the secondary observed in a numerical model by~\citet{Cuk:2010im}. In that model, the secondary of an expanding singly synchronous system is observed to begin chaotically rotating due to resonant dynamics exciting the eccentricity. They find that the secondary eventually re-synchronizes but with the opposite orientation so the secondary then evolves inward due to the BYORP effect. In their model the secondary has not evolved significantly outward when it loses synchronous rotation (all examples are within $a < 8$ $R_p$ and only evolve outward for a fraction of a $R_p$), a significantly different prediction than the currently discussed hypothesis. It is crucial to note, however, that their model has significant differences with our comprehensive model. First, there is missing a term in the second order gravity expansion, which will affect the fundamental resonances present in the system\footnote{The numerical model used in \citet{Cuk:2010im} is missing a radial acceleration term $\hat{R}$ (Section 2 of that paper). The missing term is: 
\begin{equation}
\frac{3}{2 r^4}\left( \frac{A}{2} + \frac{B}{2} - C \right)
\end{equation}
using their notation.}. Further, the model used for the BYORP effect may be too simple. The BYORP effect is a challenge to model numerically since it is very slow compared to a single orbit.~\citet{McMahon:2010by} uses averaging theory to derive the secular evolution of the mutual orbit (we use this model above). While~\citet{Cuk:2010im} make a direct, non-averaged application of Gauss' planetary equations, but to observe changes in a reasonable number of orbits they are forced to model the BYORP effect as a torque with a magnitude many times stronger than those predicted from the theory, which speeds up the evolution of the system. Lastly, the model presented above includes tides, which have been shown to be significant for the evolution of binary systems~\citep{Jacobson:2011hp}, and the YORP effect. Neither of which were considered in~\citet{Cuk:2010im}. Due to these differences, the evolutionary mechanisms in that paper seem to be fundamentally different than our current model, which considers all currently identified evolutionary effects.

\subsection{Alternative formation mechanisms}
\label{sec:alternaitve}
Right after the discovery of the first wide asynchronous binary system (1998 ST$_{27}$ by~\citet{Benner:2003ub}),~\citet{Durda:2004en} proposed a binary formation mechanism that would create wide asynchronous binaries from the debris of large impacts.  Escaping ejecta binaries (EEBs) form when two blocks of ejecta from a large impact event are on escape trajectories from the target asteroid, but the ejecta blocks are moving slowly relative to each other. If they are moving slower than their relative escape velocities and have the right angular momentum relative to their mutual center of mass, then they enter a bound orbit. According to simulations, the EEB formation mechanism can make $100$s of widely separated binaries from a single impact. These simulated binaries can match the observed separation distances of the wide asynchronous population, however EEBs are predicted to be low and high mass ratio binary systems. No high mass ratio wide asynchronous binaries are confirmed~\citep{Durda:2010vz}, 1717 Arlon may be the first (P. Pravec, personal communication). 

Most strikingly, the EEBs are not expected to have particular spin states, yet $6$ of the $9$ primaries of the wide asynchronous candidate systems have spin periods between $1$ and $2$ $P_d$  (shown in Table~\ref{tab:binaryparameters}) indicating near critical rotation rates. They are piled up at rapid rotation rates similar to the primaries of the singly synchronous binary systems~\citep{Pravec:2008cr}. Of the other three candidates: only the primary of 317 Roxane is clearly not a rapid rotator, the primary of 1717 Arlon is near rapid rotation, and the primary rotation period of 17246 (2000 GL$_{74}$) has not been measured.

These rapid primary rotation rates are consistent with formation from YORP-induced rotational fission~\citep{Margot:2002fe,Scheeres:2007io}.~\citet{Polishook:2011if} already came to a similar conclusion when they studied a sub-sample of the wide asynchronous candidates included in this study. They argue, ``The rotation periods of four out of the six objects measured by our group and others and presented here show that these suspected EEBs have fast rotation rates of 2.5 - 4 hr. Because of the small size of the components of these binary asteroids, linked with this fast spinning, we conclude that the rotational-fission mechanism, which is a result of the thermal YORP effect, is the most likely formation scenario. Moreover, scaling the YORP effect for these objects shows that its timescale is shorter than the estimated ages of the three relevant Hirayama families hosting these binary asteroids. Therefore, only the largest ($D \sim 19$ km) suspected asteroid, 317 Roxane, could be, in fact, the only known EEB.'' We note only that 317 Roxane is also consistent with the expanding singly synchronous hypothesis put forward in this paper.

\begin{figure}[tb!]
\begin{center}
\includegraphics[width=\columnwidth]{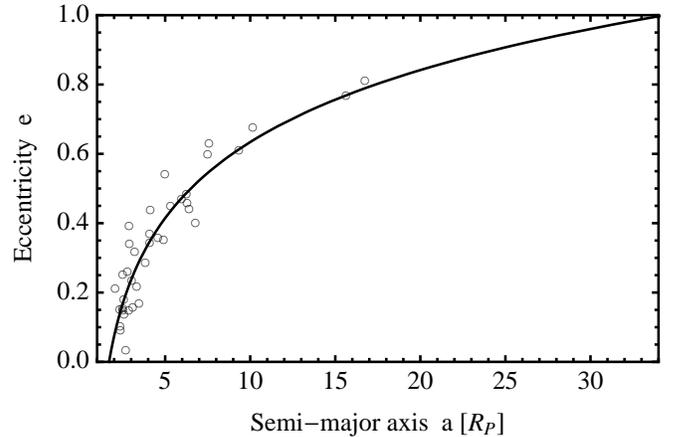}
\end{center}
\caption{The circles show the semi-major axes and eccentricities of stable binaries created by rotational fission in~\citet{Jacobson:2011eq}. The line is a simple fit to the data $e = 0.284 \ln \left( a - 0.701 \right)$.}
\label{fig:aeplot}
\end{figure}
There are two other formation mechanisms besides the EEB hypothesis that need to be addressed. First, simulations show that stable binaries can be formed at semi-major axes larger than $8$ $R_p$.~\citet{Jacobson:2011eq} found that stable binaries with semi-major axes up to $17$ $R_p$, as shown in Figure~\ref{fig:aeplot}, could form after a rotational fission event. Binary stabilization occurs after a short ($< 100$ yr) period of significant orbit transformation due to spin-orbit coupling. When binary systems do stabilize,~\citet{Jacobson:2011eq} found a relationship between the semi-major axis $a$ and the eccentricity $e = 0.284 \ln \left( a - 0.701 \right)$. The $a$-$e$ relationship is due to the near conservation of both energy and angular momentum across all three reservoirs: the orbit and the two spin states. This conservation is approximate because solar tidal perturbations on the orbit can act as either a small source or sink. Extrapolating this relationship out to an eccentricity of nearly 1, the widest binaries created directly from rotational fission are limited to obtaining a semi-major axis of $\lesssim34$ $R_p$.

Wide binary creation directly from rotational fission is rare due to this spin-orbit coupling, since increased eccentricity is often associated with further increased spin-orbit coupling leading to a positive feedback. Most systems ($\sim 92\%$) do not stabilize and eventually disrupt or collide (please read~\citet{Jacobson:2011eq} for more details). From these simulations, we can divide the wide asynchronous binaries into three groups: the widest $a > 34$ $R_p$ (1509 Escalangona, 4674 Pauling, 17246 (2000 GL$_{74}$) and 22899 (1999 TO$_{14}$) which cannot be formed directly from a rotational fission event, the tightest (1717 Arlon, 51356 (2000 RY$_{76}$) and 1998 ST$_{27}$) which are within the formation range of simulated stable binaries $a \lesssim 17 R_p$, and the intermediate $17 \lesssim a \lesssim 34$ $R_p$ (317 Roxane and 32039 (2000 JO$_{23}$)) which according to a simple fit of the simulated data could be a direct outcome of rotational fission. This intermediate range of semi-major axes correspond to very high initial eccentricities if these binaries formed directly from rotational fission.

\begin{figure}[tb!]
\begin{center}
\includegraphics[width=\columnwidth]{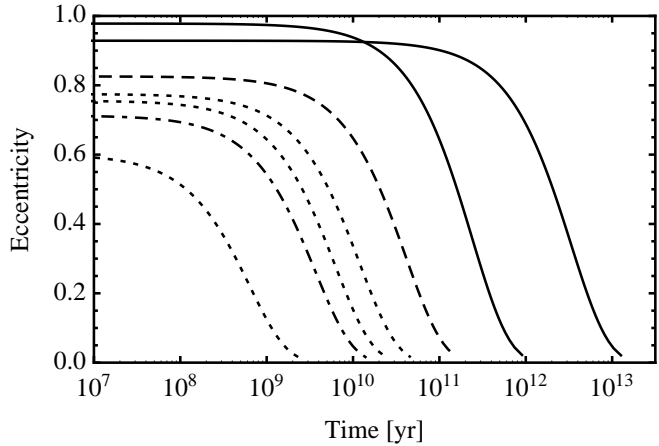}
\end{center}
\caption{Eccentricity as a function of time for each wide asynchronous candidate system. The systems from top to bottom on the left-hand side of the figure are: 32039 (2000 JO$_{23}$), 317 Roxanne, 136617 (1994 CC), 1998 ST$_{27}$, 1717 Arlon, 153591 (2001 SN$_{263}$), and 51356 (2000 RY$_{76}$). The lines are coded such that the three tightest wide binary systems (51356, Arlon and 1998 ST$_{27}$) are dotted lines, two triples systems are dashed (Balam and 136617) while 153591 is dot-dashed since it is a tight triple system, and the rest are solid black lines.}
\label{fig:eeplot}
\end{figure}
Are these initially high eccentricities subsequently damped by tides? No, and we know this from calculating the tidal evolution of each observed wide asynchronous candidate by simultaneously integrating the tidal damping equation and the tidal semi-major axis expansion equation, which are identical to Equations~\ref{eqn:eccentricity} and~\ref{eqn:semimajoraxis} but with no second, BYORP effect term. We display and discuss these equations in Section~\ref{sec:TidalandBYORPevolution} and also in Appendix~\ref{sec:parameterizationoftheevolutionequations}. Figure~\ref{fig:eeplot} shows the eccentricity damping due to mutual body tides for each of the wide asynchronous systems at their observed semi-major axes but with eccentricities given by the fitted $a$-$e$ relationship found for systems formed directly from rotational fission as shown in Figure~\ref{fig:aeplot} (i.e each model system is identical to its observed parameters except for the eccentricity). Only wide asynchronous systems with semi-major axes $a \lesssim 34 $ $R_p$ are shown, since according to the $a$-$e$ relationship, if $a \gg 34$ $R_p$, then those systems cannot form directly from rotational fission; their stable orbits would have eccentricities greater than $1$.

For each of the wide asynchronous candidates no significant tidal circularization of their orbits occurs within the relevant timescales for NEAs ($\sim10$ Myr) or MBAs ($\sim100$ Myr). Many do not damp significantly within the age of the Solar System. Since uncertainty in $k_p/Q$ translates linearly into an uncertainty in the damping timescale, if these systems are expected to damp then the assumed values from the tidal-BYORP equilibrium are incorrect by more than two orders of magnitude. Alternatively, if these systems are observed to have more circular orbits, then they did not form directly from rotational fission. The proposed expanding singly synchronous mechanism predicts more circular orbits than the direct formation mechanism and the EEB mechanism. The EEB mechanism can create circular orbits, but it preferentially creates a higher number of eccentric orbits~\citep{Durda:2004en}. Examining the mutual orbit of wide binaries will distinguish these formation mechanisms.

The last alternative formation mechanism is not so much a formation mechanism as an alteration mechanism. Planetary flybys can strongly change the mutual orbits of binary asteroids~\citep{Fang:2012go}. These flybys can expand the orbit and usually incline the orbit plane and/or excite the eccentricity. Strong changes to the mutual orbit can de-synchronize the secondary~\citep{Jacobson:2012tx}. Within the wide asynchronous candidate population, the orbit of 1998 ST$_{27}$ crosses the orbit of every terrestrial planet while the orbits of 32039 (2000 JO$_{23}$) and 51356 (2000 RY$_{76}$) barely intersect the orbit of Mars. The other $6$ wide asynchronous candidates are not planet crossing.

To summarize the three alternative formation mechanisms: (1) The ejected escaping binaries (EEB) hypothesis could produce each of these binary systems, but it would not explain why two thirds have rapidly rotating primaries consistent with formation by rotational fission. Also the mass ratio distribution does not match the predictions and all EEBs should be associated with a large asteroid family/collision. (2) Direct formation from rotational fission could produce this two thirds of the wide asynchronous candidates but it makes strong predictions regarding the eccentricity of these systems. They should be high. (3) Planetary flybys could alter already existing binary systems formed by rotational fission into wide asynchronous binary systems. Three of the nine systems have planet crossing orbits including one that crosses all four terrestrial planets. These formation mechanisms may be able to explain the existence of some of the wide asynchronous candidates, but four candidates (1509 Escalangona, 4674 Pauling, 17246 (2000 GL$_{74}$) and 22899 (1999 TO$_{14}$)) are not planet crossers, have very wide orbits, have low mass ratios and three have rapidly rotating primaries. Only the proposed expanding singly synchronous mechanism can explain these four systems and possibly the other five candidates. Overall, these formation mechanisms are not mutually exclusive and some systems could be created via one of these already proposed mechanisms.

\subsection{Individual Systems}
\label{sec:individualsystems}
If we examine each individual system in light of the four mechanisms that could produce wide asynchronous binaries--the expanding singly synchronous mechanism, escaping eject binaries~\citep{Durda:2004en}, direct formation from rotational fission~\citep{Jacobson:2011eq}, and planetary flybys~\citep{Jacobson:2012tx}--then we may be able to assess the likelihood of each mechanism for each binary. We have done some of this work already throughout the paper.

We identified in Section~\ref{sec:alternaitve}, the four systems (1509 Escalangona, 4674 Pauling, 17246 (2000 GL$_{74}$) and 22899 (1999 TO$_{14}$)) which could not form via the other three mechanisms. All three can successfully form by the expanding singly synchronous mechanism, and do so on a timescale consistent with their lifetimes. The expanding singly synchronous mechanism predicts that the mutual eccentricities of the orbits should be low although not necessarily zero. It predicts that the unmeasured primary period of 17246 (2000 GL$_{74}$) is likely rapid like the other three systems. If it is not rapid, there is the possibility that 17246 (2000 GL$_{74}$) is a doubly synchronous binary since both periods are unknown. However, it has a low mass ratio and would be the lowest mass ratio doubly synchronous system observed. Therefore, we believe it unlikely to be doubly synchronized since tidal locking of the secondary is much faster than tidal locking of the primary in low mass ratio systems~\citep{Jacobson:2011eq}. Only 1509 Exclangona has a measured asynchronous secondary. The expanding singly synchronous mechanism predicts that the other three should be asynchronous as well. 

There is the possibility that the secondaries of these three systems are synchronous. In that case, it's likely that the orbit is still expanding due to the BYORP effect. The expanding singly synchronous mechanism requires that the initial synchronization semi-major axis be within a certain range or else the system will expand past the system Hill radius. This expansion process takes time especially in the Main Belt (1s to 10s of Myrs according to Figure~\ref{fig:expsingplot}), and so discovering an expanding singly synchronous binary before it de-synchronizes or reaches the Hill radius is possible. In fact, amongst the wide asynchronous candidates all of those without measured secondary periods could be in this state including: 317 Roxane, 4674 Pauling, 17246 (2000 GL$_{74}$), 22899 (1999 TO$_{14}$) and 51356 (2000 RY$_{76}$). Alternatively systems such as 185851 (2000 DP$_{107}$), which is a synchronous system outlier when considering the tidal-BYORP effect equilibrium, may fall into this category of currently expanding singly synchronous system~\citep{Jacobson:2011hp}. Considering that the librations expected for expanding singly synchronous systems can be quite large, this motivates future observations of the libration states of synchronous secondaries. 

32039 (2000 JO$_{23}$) has a rapidly rotating primary, so it is unlikely to be an EEB, but it does barely cross the orbit of Mars, so a Martian flyby could have modified its orbit. A flyby consistent with expanding the orbit to its observed size $a \sim 32$ $R_p$ from the frequently observed tight ($a\sim4$ $R_p$) would be rare~\citep{Fang:2012go}. The system is a confirmed wide asynchronous binary and so most likely formed by the expanding singly synchronous mechanism, however it could also have formed directly from rotational fission. The test will be the eccentricity of the system. Direct formation would leave a very high eccentricity $e\sim0.98$ (see Figure~\ref{fig:aeplot}) that would not be damped in the age of the Solar System (see Figure~\ref{fig:eeplot}), while the expanding singly synchronous mechanism should produce a much lower eccentricity.

1998 ST$_{27}$ forms very quickly if it formed via the expanding singly synchronous mechanism, but its formation from this mechanism is unlikely since the YORP effect would continue to accelerate the secondary until it was rapidly rotating since it has such a small heliocentric orbit and absolute size. Since the heliocentric orbit of this system crosses the orbits of every terrestrial planet, it is more likely that this system has been excited by a planetary flyby, and is now undergoing tidally synchronization. 1998 ST$_{27}$ has an observed eccentricity $\gtrsim0.3$ that is consistent with formation from a flyby~\citep{Benner:2003ub,Fang:2012go}. Alternatively, it could have formed directly from rotational fission, but for such a wide binary, theory predicts that it should have an eccentricity $\sim 0.77$ (see Figure~\ref{fig:eeplot}). It's eccentricity should not have damped significantly (see Figure~\ref{fig:eeplot}). More precise observations of the orbit of 1998 ST$_{27}$ would determine which mechanism is more likely, since very rare and close flybys are required to create such large eccentricities~\citep{Fang:2012go}.

We find that two of the tightest of the wide asynchronous candidates (1717 Arlon and 51356 (2000 RY$_{76}$)) can only have formed via the expanding singly synchronous mechanism if the tidal ratio $X$ is within a narrow range that does not overlap much with all of the other systems (see Figure~\ref{fig:xplot}). Furthermore, in Section~\ref{sec:timescales} we find that 317 Roxane, 1717 Arlon and 51356 (2000 RY$_{76}$) require large YORP coefficients in order to rotationally accelerate to their observed rotation periods. None of these three systems cross any planetary orbits. Unlike the other two, 51356 (2000 RY$_{76}$) has a rapidly rotating primary and so if its secondary is asynchronous, it likely formed directly from rotational fission. If so, this system would initially have an eccentricity $e \sim 0.60$ and evolve according to Figure~\ref{fig:eeplot}. This system is also a good candidate for being in the expanding singly synchronous state, if the secondary is discovered to be synchronous.

317 Roxane is at the very large size end of bodies thought to be capable of YORP-induced rotational fission~\citep{Jacobson:2012vy}, and 317 Roxane and 1717 Arlon are not associated with rapid rotation. Thus neither are  likely formed by a rotational fission mechanism.~\citet{Polishook:2011if} already concluded that 317 Roxane may be an escaping ejecta binary~\citep{Durda:2004en}. 1717 Arlon may also have a high mass ratio. High mass ratio systems that start tight should be doubly synchronized due to tides, so this suggests that 1717 Arlon formed at its current wide separation. All of this evidence supports a formation by the escaping ejecta binary mechanism. Both are near the Flora family although neither~\citet{Nesvorny:2012ug} nor~\citet{Masiero:2013dk} recognize it as a family member. Despite this lack of a confirmed relationship with a large collision event, this is the only consistent mechanism.

We hesitate to comment too much about individual triple systems since the dynamics are so much richer. We wanted to carry them along in the analysis to support possible future work, but there is much to consider outside of the dynamics of just the outer pair. It is notable that~\citet{Vachier:2012jm} has identified possibly significant eccentricity in the outer pair of 3749 Balam. If so, this may rule out a simple story of binary creation, expansion, and then a second binary formation event within the outer pair. This may be an argument for a scenario as proposed in~\citet{Jacobson:2011eq}, which suggested that triple systems can be formed in a single rotational fission event, however this did not occur amongst the numerical simulations due perhaps to not enough statistics. Regardless, much more work needs to be done to understand small triple asteroid systems.

\section{Conclusions}
\label{sec:conclusions}
We developed a new formation mechanism to create the wide asynchronous binary population. A stable binary forms directly from rotational fission. The secondary synchronizes due to tides, then begins to librate. That libration is damped by tides, if the BYORP effect is expansive the system widens and an adiabatic invariance adds energy to the libration state. Later the libration ceases as the secondary begins to circulate. This turns off the BYORP effect essentially stranding the system on a wide orbit. The YORP effect rotational accelerates the secondary. These events naturally follow one another and lead to the observed properties of the wide asynchronous population.

\section{Acknowledgements}

We are grateful for the Asteroid Lightcurve Database~\citep{Warner:2009ds}, which is maintained by Brian Warner and colleagues, and the Binary Asteroid Parameters data found at \url{http://www.asu.cas.cz/asteroid/binastdata.htm}, which is compiled by methods described in~\citet{Pravec:2007fw} and maintained by Petr Pravec and colleagues. Both databases were incredibly useful for preparing this work. Seth Jacobson would also like to acknowledge the NASA Earth and Space Science Fellowship, which supported him throughout graduate school.

\appendix 

\section{Derivation of the torque ratios in Figure 1}
\label{sec:derivationofthetidaltorqueratiosinFigure1}
In Figure~\ref{fig:ratioplots}, we compare two sets of torques. In the top plot, we show the torque on the orbit from the BYORP effect divided by the torque on the orbit from the tides on the primary by the secondary:
\begin{equation}
\frac{\Gamma_B}{\Gamma_{T_o}} = \frac{H_\odot a^7}{2 \pi \rho \omega_d^2 R_p^2 q^{4/3}} \left( \frac{B_s Q}{k_p} \right)
\end{equation}
where $H_\odot = F_\odot / \left( a_\odot^2 \sqrt{1 - e_\odot^2} \right)$ is the heliocentric parameter including the solar radiation constant $F_\odot$, the semi-major axis $a_\odot$ and eccentricity $e_\odot$, $\omega_d = \sqrt{ 4 \pi \rho G / 3}$ is the surface disruption spin limit, $\rho$ is the density, $a$ is the semi-major axis measured in primary radii $R_p$, and $q$ is the mass ratio~\citep{Goldreich:2009ii,McMahon:2010jy}. The collection of tidal and BYORP effect parameters $B_s Q / k_p = 2557 (R_p / 1 \text{ km} )$ is determined by assuming the singly synchronous binary population occupies a tidal-BYORP effect equilibrium. The rest of the values are measured directly and reported in Tables~\ref{tab:binaryparameters} and~\ref{tab:tripleparameters}. The errors are dominated by the uncertainty in the tidal and BYORP effect parameters. We show two orders of magnitude for the estimated uncertainty in Figure~\ref{fig:ratioplots}.

In the bottom plot, we show the torque on the rotation state of the secondary from the YORP effect divided by the torque on the spin of the secondary from the tides on the secondary by the primary:
\begin{equation}
\frac{\Gamma_Y}{\Gamma_{T_s}} = \frac{2 H_\odot a^6}{15 \pi \rho \omega_d^2 R_p^2 q^{2/3}} \left( \frac{Y_s Q}{k_p} \right)
\end{equation}
where the YORP coefficient $Y_s$ is now a significant unknown~\citep{Scheeres:2007kv}. It is the same order of magnitude as the BYORP coefficient $\sim 10^{-2}$. We show two orders of magnitude uncertainty for the estimated ratios in Figure~\ref{fig:ratioplots}.

\section{Parameterization of the semi-major axis and eccentricity evolution equations}
\label{sec:parameterizationoftheevolutionequations}
The time evolution of the eccentricity and semi-major axis due to tides and the BYORP effect are the linear additions of each effect since they are independent of each other. The tidal equations are identical to those found in~\citet{Goldreich:2009ii} and the BYORP effect equations are identical to those found in~\citet{McMahon:2010jy}. We have re-arranged them in terms of a limited number of parameters for convenience throughout the paper. The semi-major axis and eccentricity evolution equations are:
\begin{equation}
\dot{a}  = \frac{k_p}{Q} \left( A_T +  a^7 A_B  \right) a^{-11/2} \qquad \dot{e}  = - \frac{k_p}{Q} \left( A_T L + a^7 \frac{A_B}{4}  \right) e a^{-13/2}
\end{equation}
The newly introduced parameters are:
\begin{equation}
A_T = \frac{3 \omega_d q}{ \nu^{1/2}} \qquad A_B = \frac{3 H_\odot }{2 \pi \omega_d \rho R_p^2 q^{1/3} \nu^{1/2}} \left( \frac{B_s Q}{k_p} \right) \qquad L = \frac{28 k_s  - 19 k_p q^{1/3}}{8 k_p q^{1/3} } = \frac{7 k_s}{2 k_p q^{1/3}} - \frac{19}{8}
\end{equation}
where $\omega_d = \sqrt{4 \pi \rho G / 3}$ is the surface spin disruption rate, $\rho$ is the density, $G$ is the gravitational constant, $q$ is the mass ratio, $\nu = (1+q)^{-1}$ is the fraction of the total mass in the primary, $H_\odot = F_\odot / \left( a_\odot^2 \sqrt{1 - e_\odot^2} \right)$ contains the heliocentric dependencies including the solar radiation constant $F_\odot$, the semi-major axis $a_\odot$ and eccentricity $e_\odot$. Similar to~\citet{Jacobson:2011hp}, we make the simplifying assumption that the intensive parameters of each member of a binary system are the same (e.g. density $\rho$). This is motivated by the assumption that these systems formed from rotational fission events and share a common parent body.

The BYORP coefficient $B_s$, the tidal quality number $Q$ and the tidal Love numbers $k_p$ and $k_s$ are significant unknowns. However, these formulations above conveniently collect the tidal ($k_p$ and $Q$) and BYORP ($B_s$) parameters into the coefficient $A_B$ in the same configuration as the tidal-BYORP equilibrium.~\citet{Jacobson:2011hp} determined an estimate of these parameters as a function of primary radius\footnote{This value differs slightly from that reported in~\citet{Jacobson:2011hp} since it uses updated values from the binary parameter release provided by Petr Pravec and collaborators at \url{http://www.asu.cas.cz/asteroid/binastdata.htm} using methods described in~\citet{Pravec:2007fw} and applies a Lambertian correction~\citep{McMahon:2010jy}.}: $B_s Q / k_p = 2557\ \left( R_p /1 \text{ km} \right)$. Since the BYORP coefficient $B$ is size independent and $Q$ is generally an intensive parameter,~\citet{Jacobson:2011hp} argue that $k \propto R$, which means $L = 9/8$. Using theoretical arguments,~\citet{Goldreich:2009ii} argues that $k \propto R^{-1}$ which mean $L = 7 / \left( 2 q^{2/3} \right) - 19/8$. In both cases, $L$ is positive and tides damp the eccentricity of the mutual orbit.

\section{Derivation of the adiabatic invariance between the semi-major Axis and the libration amplitude}
\label{sec:derivationoftheadiabaticinvariance}
We will use a simple model of a sphere and a triaxial ellipsoid in a mutual planar orbit (i.e. all spin and orbit poles are aligned). The primary is a sphere with radius $R_p$ and mass $M_p$. The secondary is a triaxial ellipsoid with a mass $M_s = q M_p$, where $q$ is the mass ratio. The secondary is spinning in its relaxed state about the direction of the shortest body semi-axis $\hat{z}_s$. $\hat{x}_s$ and $\hat{y}_s$ are the longest and intermediate body semi-axis directions, respectively. Thus, the moments of inertia are related $I_{s_z} = \mathsf{C} I_s > I_{s_y} = \mathsf{B} I_s  \geq I_{s_x} = \mathsf{A} I_s $ where the inertia factor of the secondary is $I_s = M_s R_s^2$. Therefore the mean radius of the secondary is $R_s  = 2.5 \left(\left( \mathsf{A} + \mathsf{B} - \mathsf{C} \right)  \left( \mathsf{B} + \mathsf{C} - \mathsf{A} \right)\left( \mathsf{C} + \mathsf{A} - \mathsf{B} \right) \right)^{1/3} = q^{1/3} R_p$. 

The coordinate tracking the instantaneous separation distance between the centers of mass of the two bodies is $r$, where $r$ and $\dot{r}$ are measured in primary radii $R_p$ and primary radii per unit time, respectively. The angle between the instantaneous line connecting the two mass centers and an arbitrary fixed line in inertial space is $\theta$. The spin angle of the $n$th body relative to the line connecting the centers is $\phi_n$.  Since the potential of the sphere is independent of its orientation, the relative spin angle of the primary sphere $\phi_p$ does not need to be tracked.

Given these definitions, the free kinetic and potential energies of the system are:
\begin{align}
T= & \nu q \frac{I_p}{2}  \left[ \dot{r}^2 + r^2 \dot{\theta}^2 + \frac{\mathsf{C} \mathcal{I}}{\nu q} \left( \dot{\theta} + \dot{\phi}_s \right)^2 \right] \\
V = & - q \frac{I_p \omega_d^2}{r} \left[ 1 -  \frac{\mathcal{I}}{4 q r^2} \left( \mathsf{A} + \mathsf{B} + \mathsf{C} -  3 \mathsf{C} \left( 1 + S \cos 2 \phi_s  \right) \right) \right]
\end{align}
\noindent where $I_p = M_p R_p^2$ is the inertia factor of the primary, $\mathcal{I} = I_s / I_p = q^{5/3}$ is the ratio of the inertia factors, $\nu = M_p / \left( M_s + M_p \right)  =  \left(1 + q \right)^{-1}$ is the fraction of the total mass in the primary,  $\omega_d = \sqrt{4 \pi \rho G / 3}$ is the surface disruption spin limit for a sphere of density $\rho$, $G$ is the gravitational constant, and $S = \left( \mathsf{B} - \mathsf{A} \right) / \mathsf{C} $ is the shape factor of the secondary. From the definition of the moments, $S=0$ for an oblate secondary, $S > 0$ for a prolate secondary, and $S$ increases with increasing prolateness of the secondary but it is limited to be less than $1$. Similar energy equations are derived in~\citet{Scheeres:2009dc}.

Using the Lagrangian $L = T - V$ and the three generalized coordinates ($r$, $\theta$ and $\phi_s$), we transform to the Hamiltonian formulation of the system. The generalized momenta for each coordinate are:
\begin{equation}
 p_r= \nu q I_p \dot{r} \qquad p_\theta =  \nu q I_p r^2  \dot{\theta} + \mathsf{C} I_s \left( \dot{\theta} + \dot{\phi}_s \right)  \qquad p_{\phi_s} =  \mathsf{C} I_s \left( \dot{\theta} + \dot{\phi}_s \right) 
\end{equation} 
The Hamiltonian $H$ is:
\begin{equation}
H =  \frac{p_{\phi_s}^2}{2 \mathsf{C} I_s} + \frac{1}{2 \nu q I_p} \left[ p_r^2 + \frac{ \left(p_\theta - p_{\phi_s} \right)^2}{q_r^2} \right] + V
\end{equation}
The instantaneous equations of motion for the system can be determined from this Hamiltonian for the generalized coordinates given above, but we can reduce the number of canonical pairs by introducing an integral of motion. From the equations above, it is clear that the coordinate $\theta$ is ignorable: $\partial L / \partial \theta = 0$. This is the conservation of angular momentum, and the conserved quantity is:
\begin{equation}
K = \frac{\partial L } {\partial \dot{\theta}} = \nu q I_p r^2 \dot{\theta} +  \mathsf{C} I_s \left( \dot{\phi}_s + \dot{\theta} \right) = p_\theta
\end{equation}
The generalized momentum for the relative spin angle of the secondary $p_{\phi_s}$ can then be expressed solely in terms of its coordinate:
\begin{equation}
p_{\phi_s} = \mathsf{C} I_s \dot{\phi}_s \left[ 1 +  \frac{ \mathsf{C} \mathcal{I} }{\nu q  r^2} \right]^{-1} 
\end{equation} 

Considering the system that we wish to study, the changes in the instantaneous separation distance $\delta r$ are very small compared to the instantaneous separation distance $r$, so $\delta r \ll r$. We make the approximation that $\dot{r} \approx 0$. This implies that the orbit is circular, so $r \approx a$, where $a$ is the semi-major axis measured in primary radii. The semi-major axis will change over time due to mutual body tides and the BYORP effect, but that change is very slow compared to the orbit or libration periods, so we assume $\dot{r} = 0$. Therefore in the Hamiltonian system, the generalized momenta $p_r = 0$. Furthermore when $\delta r \ll r$, the instantaneous rotation of the line connecting the two mass centers relative to inertial space is then the mean motion $\dot{\theta} = n$. This approximation means that Kepler's third law is valid: $\nu a^3 n^2 = \omega_d^2 $.

Using a collection of constant terms, the single degree of freedom Hamiltonian is:
\begin{equation}
H = H_0 + H_1 p_{\phi_s}^2 - H_2 \cos 2 \phi_s
\end{equation}
where
\begin{align}
H_0 = & - q \frac{I_p \omega_d^2}{a} \left[ 1 - \frac{\mathcal{I}}{4 q a^2} \left( \mathsf{A} + \mathsf{B} - 2 \mathsf{C} \right) \right]  - \frac{K^2}{2 \nu q I_p a^2}  \left( 1 + s  \right)^{-1} \\
H_1 = & \frac{1+s}{2 \mathsf{C} I_s} \\
H_2 = &  \frac{3 S \mathsf{C} I_s \omega_d^2}{4 a^3}
\end{align}
where $s =  \mathsf{C} \mathcal{I} / \nu q a^2$ is the secondary perturbation term. It appears throughout the rest of these derivations and in many of the final expressions. This quantity approaches zero as the semi-major axis increases, the mass ratio decreases, and the secondary maximum moment of inertia decreases relative to the primary moment of inertia factor.

The Hamiltonian equations of motion are:
\begin{equation}
\dot{p}_{\phi_s} =  - 2 H_2 \sin 2 \phi_s \qquad \dot{\phi}_s =  2 H_1 p_{\phi_s}
\label{eqn:hamiltonianequationsofmotion}
\end{equation}
Let's identify some features of this system. First, there are stable equilibria at $\phi_s = 0$, $\pm \pi$ and $p_{\phi_s} = 0$. This corresponds to the long axis relative equilibrium described in~\citet{Bellerose:2008fo}. There are also unstable equilibria at $ \phi_s = \pm \pi / 2$ and $p_{\phi_s} =  0$, which correspond to the short axis relative equilibria in~\citet{Bellerose:2008fo}. Second, there is a separatrix that divides the motion of the secondary between libration and circulation. This separatrix goes through the unstable equilibria. The value of the Hamiltonian at the separatrix is:
\begin{equation}
H_S = H_0 + H_2
\end{equation}
The secondary is on the separatrix if $H = H_S$, librating if $H < H_S$, and circulating if $H > H_S$. Third, when the system is librating $H < H_S$, the Hamiltonian can be expressed in terms of the libration amplitude of the secondary $\Phi_s$ as it librates about $\phi_s = 0$. When the secondary is at the maximum libration angle $\phi_s = \Phi_s$, the relative spin velocity is $\dot{\phi}_s = 0$ so the system is on the separatrix. Therefore, the Hamiltonian can also be expressed as:
\begin{equation}
H = H_0  - H_2 \cos 2 \Phi_s = H_S - 2 H_2 \cos^2 \Phi_s
\end{equation}
The second expression of the Hamiltonian clearly shows that the system is librating as long as $|\Phi_s| < \pi /2$. When $|\Phi_s| = \pi/2$, the system is on the separatrix. This paper explores this dynamical system in this librating regime under the influence of orbit expansion and tides.

The action of the relative spin angle of the secondary is the integrated phase space for a full cycle of the coordinate $\phi_s$: $J_{\phi_s} = \oint p_{\phi_s}\ d q_{\phi_s} $. Since we are considering very slow orbit expansion due to the BYORP effect and tides on the primary and slow libration damping of the secondary relative to both the libration and orbital periods, the action is an adiabatic invariant. Using the equations above, the conjugate momentum of the relative spin angle of the secondary $p_{\phi_s}$ can also be expressed as:
\begin{equation}
p_{\phi_s} =  \sqrt{ \frac{H_2}{H_1}  }  \sqrt{\cos 2 \phi_s - \cos 2 \Phi_s}
\label{eqn:reexpresshamiltonian}
\end{equation}
The action can be integrated directly using the expression for the conjugate momentum above but the result is a function of an incomplete elliptical integral of the second kind. It is more useful to transform the action by substituting in the variable $\chi$ using the trigonometric relation $\sin^2 \chi = \sin^2 \phi_s / \sin^2 \Phi_s$: 
\begin{equation}
J_{\phi_s} = \sqrt{ \frac{8 H_2 }{H_1 }  } \int_{-\Phi_s}^{\Phi_s}  \sqrt{ \sin^2 \Phi_s \left[1 - \frac{\sin^2 \phi_s}{\sin^2 \Phi_s} \right] }\ d\phi_s = \sqrt{ \frac{32 H_2 }{H_1} }  \int_{0}^{\frac{\pi}{2}} \frac{\sin^2 \Phi_s \cos^2 \chi}{ \sqrt{ 1 -\sin^2 \Phi_s  \sin^2 \chi  }}\ d\chi
\end{equation}
Now, we integrate the action using complete elliptic integrals:
\begin{equation}
J_{\phi_s} = \frac{4 \mathsf{C} I_s \omega_d }{a^{3/2}} \mathsf{G}(\sin^2 \Phi_s) \sqrt{\frac{3 S}{1+s}} 
\label{eqn:adiabaticinvariance}
\end{equation}
The generalized solution $\mathsf{G}(k^2)$ to this integral can be expressed in terms of complete elliptic functions of the first $\mathsf{K}(k^2)$ and second $\mathsf{E}(k^2)$ kind:
\begin{equation}
 \mathsf{G}(k^2) = \int_0^{\pi/2} \frac{ k^2 \cos^2 x}{\sqrt{1 - k^2 \sin^2 x}}\ dx =  \mathsf{E}(k^2) - \left( 1 - k^2 \right) \mathsf{K}(k^2) 
\end{equation}

\section{Derivation of the libration growth due to orbit expansion}
\label{sec:Derivationofthelibrationgrowthduetoorbitexpansion}
In order to get a relationship between the time derivative of the semi-major axis and the libration amplitude, we take the time derivative of the square of the adiabatic invariance (Equation~\ref{eqn:adiabaticinvariance}):
\begin{equation}
\frac{\mathsf{K}(\sin^2 \Phi_s)  \sin 2 \Phi_s}{\mathsf{G}(\sin^2 \Phi_s) } \dot{\Phi}_s - \frac{ 3 + s}{1+s} \left( \frac{\dot{a}}{a}\right) =0 
\end{equation}
The orbit expansion $\dot{a}$ is due to both tides and the BYORP effect:
\begin{equation}
\frac{\dot{a}}{a} = \frac{3 k_p \omega_d q }{Q_p a^{13/2} \nu^{1/2}} + \frac{3 H_\odot B_s a^{1/2} }{2 \pi \omega_d \rho R_p^2 q^{1/3} \nu^{1/2}}
\end{equation}
It is important to note that this tidal orbit expansion is not due to the libration tides on the secondary, but the circulation tides on the primary. We assume a primary rotating prograde and rapidly (i.e. rotation rate faster than the mean motion). 

From here it is simple to state the time derivative of the libration amplitude due to BYORP and tidal driven orbit expansion:
\begin{equation}
\dot{\Phi}_s =  \frac{ 3 + s}{1+s}  \left[ \frac{3 k_p \omega_d q }{Q_p a^{13/2} \nu^{1/2}} + \frac{3 H_\odot B_s a^{1/2} }{2 \pi \omega_d \rho R_p^2 q^{1/3} \nu^{1/2}} \right] \frac{\mathsf{G}(\sin^2 \Phi_s) }{\mathsf{K}(\sin^2 \Phi_s)  \sin 2 \Phi_s} 
\end{equation}

\section{Derivation of the energy dissipation rate due to libration tides}
\label{sec:derivationoftheenergydissipationrateduetolibrationtides}

We are searching for the energy dissipated within the rotation state of a binary asteroid member due to mutual body tides.  To simplify this problem, we will de-couple the system and, for the determination of the tidal energy dissipation only, treat each body as a sphere.\footnote{Arbitrary shapes could be treated for either the potential of the tide raising body if it is expanded in spherical harmonics or the surface of the secondary if it is expanded in surface spherical harmonics. This may be future work.} This derivation follows those given in~\citet{Wisdom:2004dy,Wisdom:2008gr}, and we repeat material to inform the reader. Wisdom derives tidal dissipation induced by various forced librations and non-zero obliquities, we derive the tides for a free libration in a circular orbit below. Deriving these free libration tides for an eccentric orbit requires more than one set of tidal responses due to the multiple, non-harmonic forcing frequencies and is left for future work. 

The energy dissipated within the interior of a homogenous (constant density) body moving through the gravity field of a point mass is the work done on each individual element within the body. The work  can be expressed as the dot product of the tidal force on that element of the body $\vec{F}_T$ with the displacement of the element $\delta \vec{x}$, which is also the instantaneous velocity $\vec{v}$ of the element over an instant of time $\delta t$: $\delta W_T = \vec{F}_T \cdot \delta \vec{x} = \vec{F}_T \cdot \vec{v}\ \delta t $, which we rearrange to express as a rate of work done per element. 

The tidal force $F_T$ is a negative gradient of the perturbing potential energy $V_T$, which is the perturbing tidal potential $U_T$ multiplied by the mass of the element $dm = \rho dV$. The expression for the rate of work done on each element can be integrated over all volume elements of the body to determine the total rate of change in energy.
\begin{equation}
\dot{E}_T = \iiint\limits_\text{Body} \dot{W}_T\ dV = - \rho \iiint\limits_\text{Body} \vec{v} \cdot \nabla U_T dV
\end{equation}
Assuming that the body is incompressible $\nabla \cdot \vec{v} = 0$, we make a simple substitution using the product rule: $ \nabla \left( U_T \vec{v} \right) = \vec{v} \cdot \nabla U_T $, and use Gauss's theorem to express this volume integral as a surface integral:
\begin{equation}
\dot{E} =  - \rho \iiint\limits_\text{Body} \nabla \left( U_T \vec{v} \right) dV = - \rho \iint\limits_\text{Body} U_T \vec{v} \cdot \vec{n}\ dS
\end{equation}
where $dS$ is the area of the particular surface element and $\vec{n}$ is the unit normal direction to that surface. 

Love (1948) determined that the radial displacement height of a surface element $\Delta r = - h U_T' / g$ is linearly dependent on the displacement Love number $h = 5/3 k$, which is related to the potential Love number $k$, and the delayed tidal potential $U_T'$ (delayed because the dissipative response lags the forcing), and inversely dependent on the surface acceleration due to gravity $g$. The time rate of change of this displacement is conveniently $\vec{v} \cdot \vec{n}$. The energy dissipation can now be expressed directly as the response of the surface of the body to the tidal potential and its time derivative:
\begin{equation}
\dot{E} = \frac{\rho h}{g} \iint\limits_\text{Body} U_T \frac{d}{d t} \left( U_T' \right)\ dS
\label{eqn:energydissipationwithintegrals}
\end{equation}

For the problem at hand, the dissipation occurs in the secondary so the tidal raising perturbing potential  is that of the primary:
\begin{equation}
U_T = - \frac{\omega_d^2 R_s^2}{a^3} P_2 \left( \cos \alpha \right)
\end{equation}
where $P_2 \left( \cos \alpha \right)$ is the second Legendre polynomial, and $\alpha$ is the angle at the center of the second perturbed body between the vector from the center of the second body to the first $\vec{o}$ and the vector from the center of the second body to the surface element $\vec{s}$. The length of each vector is known: $|\vec{o}| = R_s$ and $|\vec{s}| = a$, and we determine $\cos \alpha = \vec{o} \cdot \vec{s} / (a R_s)$. The vector from the center of the secondary to the center of the primary $\vec{o}$ in cartesian coordinates is, for a circular orbit: $ \vec{o} = ( a \cos n t, a \sin n t, 0 ) $ with mean motion $n$.

We decompose the vector from the center of the secondary to a surface element $\vec{s}$ into the motion of the surface of the body relative to the center of the secondary $\mathcal{R}$ (i.e. some rotation matrix) and a vector from the center of the secondary to a surface element described by a planetocentric longitude $\lambda$ and colatitude $\theta$ at some initial time, in cartesian coordinates: $\vec{s} = \mathcal{R} \vec{s}_0 =  \mathcal{R} ( R_s \sin \theta \cos \lambda, R_s \sin \theta \sin \lambda, R_s \cos \theta )$. 

We can consider a number of surface motions that the second body could be making, but for now we only consider free libration of the secondary in a circular mutual orbit. The body is rotating at the rate of the mean anomaly $n t$ and librating with a frequency $\omega_l$ and an amplitude $\Phi_s$. The surface rotation matrix $\mathcal{R}$ is:
\begin{equation}
\mathcal{R} = \left( \begin{array}{ccc}
\cos \left( n t - \Phi_s \sin \omega_l t \right) & - \sin \left( n t - \Phi_s \sin \omega_l t \right) & 0 \\
\sin \left( n t - \Phi_s \sin \omega_l t \right) & \cos \left( n t - \Phi_s \sin \omega_l t \right) & 0 \\
0 & 0 & 1
 \end{array} \right)
\end{equation}
where $\omega_l$ is the libration frequency as before. We expand the rotation matrix to first order in the libration angle amplitude $\Phi_s$ and from these equations, we determine the tidal potential $U_T$:
\begin{align}
U_T =  \frac{\omega_d^2 R_p^2 q^{2/3}}{16 a^3} \bigg[  & 2 - 3 \Phi_s^2 + 3 \left( 2 + \Phi_s^2 \right) \cos 2 \theta + \nonumber \\
& 6 \sin^2 \theta \left( \left( \Phi_s^2 -2 \right) \cos 2 \lambda + 2 \Phi_s \left( \Phi_s \sin^2 \lambda \cos 2 \omega_l t - 2 \sin 2 \lambda \sin \omega_l t \right) \right) \bigg]
\label{Eq:tidalpotential}
\end{align}

In order to determine the delayed potential, ~\citet{Wisdom:2004dy} says, ``The delayed potential $U_T'$ is found by replacing $n t$ by $ nt + \Delta$ in the expression for $U_T$.'' This is equivalent to making the substitution $t \longrightarrow t + \Delta / n $. For the systems considered in~\citet{Wisdom:2004dy,Wisdom:2008gr}, the tidal forcing frequency is always $n$ or a rational factor of $n$ (e.g. $1/3$). Since we are considering a freely librating body in a circular orbit, the tidal forcing frequency is the libration frequency $\omega_l$. Thus, an appropriate\footnote{If we were considering an eccentric orbit, there would be two fundamental forcing frequencies: $\omega_l$ and $n$, which would necessitate a separate treatment for each, since each excitation would have it's own tidal phase lag ($\Delta_l$ and $\Delta$) and tidal dissipation number ($Q_l$ and $Q$).} substitution rule is $t \longrightarrow t + \Delta_l / \omega_l$. Here, we have distinguished the tidal phase lags: $\Delta_l$ and $\Delta$, since these lags correspond to the libration frequency and mean motion forcing frequencies, respectively. Likewise, we will distinguish the tidal quality numbers: $Q_l$ and $Q$, which correspond to each forcing frequency.

After determining $U_T'$, we return to Equation~\ref{eqn:energydissipationwithintegrals} and perform the surface integral on the secondary:
\begin{equation}
\dot{E} = - \frac{2 \pi k_s \rho \omega_d^2 \omega_l \Phi_s^2 R_p^5 q^{5/3}}{3 a^6} \left( 3 \sin \Delta_l + \sin \left( 2 \Delta_l + 2 \omega_l t \right) - 3 \sin \left( \Delta_l + 2 \omega_l t \right) \right)
\end{equation}
This is the instantaneous energy dissipation rate of free libration within the secondary due to tides.

Similar to \citet{Wisdom:2004dy,Wisdom:2008gr}, we can remove the time dependance of this equation by averaging\footnote{This averaging is acceptable as long as the orbital period is much shorter than the energy dissipation timescale, which is always the case here.} over a forcing (libration) period.
\begin{equation}
\dot{E}  =  - \frac{2 \pi k_s \rho \omega_d^2 \omega_l \Phi_s^2 R_p^5 q^{5/3}}{ Q_l a^6}
\label{eqn:tidalenergydissipation}
\end{equation}
where we removed the tidal lag angle dependance using the canonical relationship $\sin \Delta_l = Q_l^{-1}$. The tidal quality number is defined: $Q_l =  2 \pi E_0 / \oint \dot{E}\ dt$ where $E_0$ is the peak energy stored in the system during a forcing cycle and $\oint \dot{E}\ dt$ is the energy dissipated over the cycle. 

\paragraph{Note} This theory shares the same difficulty as the~\citet{Mignard:1979kb,Mignard:1980bd} model where the tidal bulge could potentially wrap around the body. We can determine a condition for this to occur, and make sure that we are safely outside of those bounds. If the maximum tidal bulge angle is $\Delta_l < \pi/2 $ before wrapping occurs, then this constrains the tidal quality number to $Q_l >1$, which implies that the system must be underdamped in order for the bulge not to wrap.  Advantageously, this relationship between the tidal forcing and response avoids the awkward tidal switching that occurs with a constant tidal lag angle model (e.g.~\citep{Goldreich:1963wa}). This discontinuity, which occurs when the tidal bulge angle (i.e. the spin angle $\phi_s$) goes through zero, causes the tidal bulge to jump across the body.

\section{Derivation of the libration frequency}
\label{sec:derivationofthelibrationfrequency}
We derive the libration frequency $\omega_l$ from the definition of the libration period $T = 2 \pi / \omega_l = \oint \dot{\phi}_s^{-1}\ d \phi_s $ where $\phi_s$ is the spin angle of the secondary. When $\phi_s = 0$, the longest axis $\hat{x}_s$ of the secondary is aligned with the line connecting the mass centers of the two bodies. Using the Hamiltonian equations of motion (\ref{eqn:hamiltonianequationsofmotion}) and the Hamiltonian written as a function of the libration angle amplitude $\Phi_s$ (\ref{eqn:reexpresshamiltonian}), the libration period is:
\begin{equation}
T = \sqrt{ \frac{2}{H_1 H_2}} \int_0^{\Phi_s} \frac{1}{\sin \Phi_s} \left[ 1 - \frac{\sin^2 \phi_s}{\sin^2 \Phi_s} \right]^{-\frac{1}{2}}\ d\phi_s = \sqrt{ \frac{2}{H_1 H_2}} \mathsf{K} \left( \sin^2 \Phi_s \right)
\end{equation}
where $K(k^2)$ is the complete elliptic function of the first kind. The integration is done directly after substituting the variable $\chi$ using the trigonometric relation $\sin^2 \chi = \sin^2 \phi_s / \sin^2 \Phi_s$. 

The libration frequency $\omega_l$ is:
\begin{equation}
\omega_l = \frac{\pi \sqrt{ 2  H_1 H_2}}{ \mathsf{K}(\sin^2 \Phi_s) }= \frac{ \pi \omega_d \sqrt{ 3 S \left( 1+ s \right)} }{  2 \mathsf{K} (\sin^2 \Phi_s) a^{3/2} } 
\end{equation}
For low mass ratio systems $q \ll 1$, the libration frequency is proportional to the mean motion: $\omega_l \approx \pi n \sqrt{3 S} / 2 \mathsf{K}(\sin^2 \Phi_s)$. For small libration angle amplitudes $\Phi_s \ll 1$, the relationship is merely a function of the shape of the secondary: $\omega_l \approx \pi n \sqrt{3 S / 4 }$.

\section{Derivation of the Libration Damping Due to Tides}
\label{sec:derivationofthelibrationdampingduetotides}
Using the model described in Appendix~\ref{sec:derivationoftheadiabaticinvariance}, the energy of the system is the Hamiltonian: $ E = H_0 - H_2 \cos 2 \Phi_s$, which is described solely as a function of the libration amplitude $\Phi_s$. It is important to remember that  due to their symmetry the dissipation of libration tides does not  secularly evolve the mutual orbit. We take the time derivative of the energy:
\begin{equation}
\dot{E} = \frac{3 S \mathsf{C} I_s \omega_d^2}{2 a^3} \dot{\Phi}_s \sin 2 \Phi_s 
\end{equation}
where $S =  \mathsf{B} - \mathsf{A} / \mathsf{C} $ is a shape factor. We set the time derivate of energy equal to the energy dissipation result found at the end of Appendix~\ref{sec:derivationoftheenergydissipationrateduetolibrationtides} in equation \ref{eqn:tidalenergydissipation}. Thus,  the time derivative of the libration amplitude due to tidal damping is:
\begin{equation}
\dot{\Phi}_s = - \frac{k_s \omega_l \Phi_s^2}{ Q_l  S a^3 \sin 2 \Phi_s  }
\label{eqn:tidaldamping}
\end{equation}

\bibliographystyle{apj}
\bibliography{bibliography.bib}

\end{document}